\documentclass[12pt, aps, nofootinbib]{revtex4-1}
\usepackage{amsmath,amssymb}
\usepackage{verbatim,graphicx}

\newcommand{\ket}[1]{| #1\rangle}
\newcommand{\bra}[1]{\langle #1|}
\newcommand{\braket}[2]{\langle #1|#2\rangle}

\newcommand{\BZ}{\mathbb{Z}}

\def\Tr{\textrm{Tr}}
\def\lp{\left(}
\def\rp{\right)}
\def\lb{\left[}
\def\rb{\right]}

\def\lr{\left.}
\def\rr{\right.}

\newcommand{\beq}{\begin{equation}}
\newcommand{\beqs}{\begin{equation*}}
\newcommand{\eeq}{\end{equation}}
\newcommand{\eeqs}{\end{equation*}}

\allowdisplaybreaks

\begin{document}
\setlength{\unitlength}{1mm}
\title{Open spin chains for giant gravitons and relativity}

\author{David Berenstein, Eric Dzienkowski}
\affiliation{Department of Physics, University of California at Santa Barbara, CA 93106}

\begin{abstract} We study open spin chains for strings stretched between giant graviton states in the ${\cal N}=4$ SYM field theory in the collective coordinate approach.  We study the boundary conditions and the effective Hamiltonian of the corresponding spin chain to two loop order. 
  The ground states of the spin chain have energies that match the relativistic dispersion relation characteristic of massive $W$ boson particles on the worldvolume of the giant graviton 
configurations,  up to second order in the limit where the momentum is much larger than the mass. We  find evidence for a non-renormalization theorem for the ground state wave function of this spin chain system.  We also conjecture a generalization of this result to all loop orders which makes it compatible with a fully relativistic dispersion relation. We show that the conjecture follows if one assumes that the spin chain admits a central charge extension that is sourced by the giant gravitons, generalizing the giant magnon dispersion relation for closed string excitations.  
This provides evidence for ten dimensional local physics mixing AdS directions and the five-sphere emerging from an ${\cal N}=4$ SYM  computation in the presence of a non-trivial background (made of D-branes) that break the conformal field theory of the system. \end{abstract}

\maketitle

\section{Introduction}
\label{sec:Introduction}

The AdS/CFT correspondence \cite{Maldacena:1997re} suggests that there should be a relation between the low energy effective field theories of string theory (both with open and closed strings) or M-theory on AdS spaces and the dual gauge theory. These effective field theories in higher dimensions are local and locally Lorentz invariant in the supergravity limit.
One of the biggest puzzles in understanding the AdS/CFT correspondence is on exactly how locality in higher dimensions emerges from lower dimensional physics. One should argue that if this locality and higher dimensional geometry is emergent, locality should be found at the end of some computation. There are various ways of trying to get a handle on this problem.

A natural way to express the emergence of a local theory is that one should be able to reproduce the flat space S-matrix in the appropriate limit. Some progress in this direction was made in \cite{Gary:2009ae, Fitzpatrick:2011ia}, but it has also been argued that sub-AdS holography might fail \cite{Giddings:1999jq} and in spite of the new developments one still has to worry about this possible failure  because of tail properties of wave packets that are used to probe scattering \cite{Gary:2009mi}.
It is expected that in order for local physics to work,  a gap in anomalous dimensions develops \cite{Heemskerk:2009pn} and that one can count constraints arising from crossing symmetry in correlators and match them to polynomial terms in the AdS effective action. With enough assumptions about locality in the bulk one can develop a picture where CFT perturbation theory might be attempted \cite{Fitzpatrick:2010zm}. But this is very different than predicting locality from a CFT computation. Instead, one can argue that this approach  shows that effective field theory in the bulk makes sense from the point of view of field theory constructions on the boundary. This setup has not furnished a proof of locality. It also doesn't explain what happens when the conformal symmetry of the bulk is broken, and moreover, local physics on $AdS_5\times S^5$ is not the same as local physics on $AdS$: one can have scattering into the sphere directions from wave packets starting in AdS. Thus, one needs to show that physics is also local along the sphere directions and that indeed, one should have local Lorentz symmetry in ten dimensions, not five.

There are other features present in the $AdS_5\times S^5$ geometry that can help with addressing the problem of locality, or at least, local Lorentz invariance. This is the fact that the string motion of strings propagating on $AdS_5\times S^5$ seems to be integrable.  It has been  argued that integrability shows that local strings moving in $AdS\times S$ are related to certain planar calculations on the boundary, where a similar integrable structure has been found. Thus, one should be able to see how the dynamics of the higher dimensions can be recovered from a field theory computation.  For a review of integrability see \cite{Beisert:2010jr}. Integrability is very special and it does not persist in more general setups like general marginal deformations of $N=4$ SYM \cite{Berenstein:2004ys} or the  conformal field theory of the conifold where it is known that the string sigma model is not integrable \cite{Basu:2011di}. It also does not explain what happens when conformal invariance is broken. The computation of scattering between string states in the integrability program is still a work in progress. 

One can ask if there is an alternative way to approach this problem of locality. Indeed, some partial progress towards understanding how locality arises (along the sphere $S^5$ of $AdS_5\times S^5$) can be made 
by guessing which field configurations could dominate the strong coupling physics \cite{Berenstein:2005aa}, at least in the BPS case, and finding an approximation for the perturbative spectrum around such configurations.
At the moment this is an uncontrolled approximation, but it has the advantage that the conformal symmetry of the theory plays a smaller role in this setup, and that within this approach one can produce different backgrounds for the gravity theory (at least in principle) by expanding around different BPS configurations. 

Because probing local physics depends on being able to localize particles on short distances, one can imagine that having objects with a shorter Compton wavelength (more massive particles) make better probes of the local geometry
those that have a longer Compton wavelength .
A successful strategy to tackle the locality problem  might start by studying objects that are not strings first, but much heavier objects. Such objects could not be objects in the low energy effective field theory of the AdS supergravity either, as all of the supergravity fluctuations are string states. One can therefore argue that non-perturbative stringy effects might be more amenable to study geometry than supergravity fields. The natural candidates for such objects are D-branes (and membranes), which have also well defined local effective actions describing their motions in the geometry.  Thus, they might furnish a well defined local probe of the geometry that can be studied in detail. In order to find such states in the field theory, one needs to be able to control them at weak coupling. Ideally they are BPS states that one can follow from weak coupling to strong coupling. This paper deals with an example of such states.

Giant gravitons \cite{McGreevy:2000cw} and their cousin D-brane excitations of AdS geometries have become a very useful tool to understand geometric aspects of the AdS/CFT duality and thus are a natural test of this idea.  They also provide in general BPS objects whose quantum numbers (and some of their BPS excitations) are calculable and whose degeneracies can be compared with field theory data \cite{Witten:1998xy,Gubser:1998fp,Mikhailov:2000ya,Berenstein:2002ke}. One can start examining the local physics with one such D-brane probe, but the physics is much richer if one has multiple D-brane probes of the geometry. Such a setup can help us distinguish {\em here from there}, exactly because the first D-brane can be placed here and the second one can be placed there. Given the two D-branes, we can measure the distance from here to there by stretching a string between them and computing its energy. Thus the energies of strings start serving as a probe of the metric of the geometry. As we change the coupling constant in the dual CFT, the string tension changes in AdS units, and distances of sub-AdS regions can become very large in string units. One can use this to study how the gap between excitations develops and thus how these massive string excitations can eventually decouple in various processes. Moreover, such strings can move along the D-branes (if they are extended), so one can also test if they are compatible with locally Lorentz invariant  physics in ten dimensions, or more precisely, along the world-volume of the D-brane. This is, one should find that they are relativistic. In the field theory limit the low lying strings should have a relativistic dispersion relation.  This is usually taken for granted in the AdS geometry. However, this is not granted from the beginning of the calculation in the dual field theory. One can moreover argue that the presence of the D-branes breaks conformal invariance, so that  dispersion relations in the presence of a D-brane is not just kinematics of the AdS isometry group. It is instead a fully dynamical derived quantity.

From the point of view of semiclassical calculations in gravity, we usually think of these D-branes as being perfectly localized in the AdS geometry. However, from the point of the dual field theory, most of these excitations are delocalized, because when we classify them we usually require that they have precise R-charge quantum numbers and therefore fixed energy. The gauge/gravity duality is a quantum equivalence between two formulations and as such is subject to the uncertainty principle of quantum mechanics. Indeed, the dual operators to the giant gravitons with fixed R-charge have been identified \cite{Balasubramanian:2001nh}. Since the R-charges are usually a momentum quantum number in the geometry, the uncertainty principle requires that the D-branes be fully delocalized in the dual geometric variable, which is an angle in the geometry.

To address the problem of the position of the brane in the AdS geometry directly from field theory, one needs to find a collective coordinate that parametrizes generalized coherent states which permit a geometric interpretation of the position of the D-brane as a specific value of this collective coordinate. Given two such D-brane excitations of the geometry we can then in principle start studying the spectrum of strings stretching between them and in this way obtain metric information about the geometry in which they live. Without localized D-branes, in principle the naive spectrum of strings stretching between them would average over configurations and would lose the geometric meaning. One can argue that diagonalizing the effective Hamiltonian for stretched strings between the D-branes would restore the collective coordinates (or more precisely the relative collective coordinates) at the expense of solving a complicated mixing problem in degenerate perturbation theory. Such a program can be carried out in some setups (see \cite{deMelloKoch:2011ci} and references therein). This is simplified considerably if one begins with a tractable collective coordinate formulation {\it ab initio} where there is no mixing, or more precisely, the mixing is very suppressed between different choices of the values of the collective coordinate. That is, one can argue that the effective Hamiltonian of strings stretching between branes is local in the effective coordinate\footnote{One can consider back-reaction effects of the D-branes - in particular recoil - as $1/N$ corrections, see for example \cite{Berenstein:1996xk}, and this should be neglected on a first pass.}.

One of us recently introduced such a collective coordinate approach to study giant gravitons with their excitations \cite{Berenstein:2013md}. The technique introduced a complex collective coordinate for the giant graviton state in ${\cal N} = 4$ SYM field theory that is restricted to live on a disk of radius $\sqrt{N}$ centered around the origin. That coordinate has a clear geometric interpretation in terms of the fermion droplet description of half BPS states \cite{Berenstein:2004kk}, which was later understood to describe the general half-BPS geometry with $AdS_5\times S^5$ boundary conditions \cite{LLM}.

The present paper utilizes this new collective coordinate approach to compute the open spin chain boundary conditions for strings stretched between giant gravitons in the $SU(2)$ sector, generalizing previous results \cite{Berenstein:2005vf,Berenstein:2005fa,deMelloetal,Berenstein:2007zf}. The most important improvement factor is that we do computations in the presence of the boundary to two loop order for a string stretched between two different branes. 

The paper is organized as follows. In section \ref{sec:cuntz}, we set up the problem of studying the one loop Hamiltonian for the $SU(2)$ spin chain. 
We review how the $SU(2)$ spin chain that is obtained has in principle a variable number of sites, but after a bosonization transformation it can be regarded as a bosonic spin chain with a fixed number of sites, with nearest neighbor hopping, and non-diagonal boundary conditions \cite{Berenstein:2005fa}. The natural bosonic degrees of freedom are described by a Cuntz oscillator at each site, and we also describe the boundary conditions that are induced by the giant graviton states in detail.
In section \ref{sec:coh} we study the ground state of this spin chain by using coherent states for the Cuntz oscillator, which are the natural degrees of freedom for this spin chain and we also compute the ground state energy. The Cuntz oscillator coherent states are also described by a complex parameter and are restricted to a disk of radius $1$. We show that the coordinates of the disk of radius one arising from the spin chain and the disk of radius $\sqrt{N}$ are really describing the same disk after rescaling and complex conjugation. However, the effective geometry that the string and the giant graviton see are different from each other. This can be understood in terms of a Berry phase contribution to a first order effective action describing the dynamics.
Next, in section \ref{sec:2-loop} we study the spin chain Hamiltonian up to two loop order. It is important for us how the local spin chain Hamiltonian looks like in terms of the Cuntz oscillators. We are able to show that the ground state of the one loop Hamiltonian is still an eigenfunction of the two loop Hamiltonian. We can also compute the two loop energy exactly and find that some terms in the effective Hamiltonian vanish exactly for these states. 
In section \ref{sec:lorentz} we see that for these ground states we are starting to build a relativistic dispersion relation to second order, where the momentum is related to the number of sites on the spin chain, and the effective mass is related to the boundary conditions and depends on the gauge coupling constant. We generalize from these observations what the higher loop result should look like to all orders. We also show that the associated relativistic dispersion relation should be a consequence of the central charge extension of the integrable spin chain setup \cite{Beisert:2005tm}, generalized to this open string sector. Then we  conclude.

\section{The Cuntz spin chain boundary conditions}
\label{sec:cuntz}

Let us first briefly recall the determination of the one loop anomalous dimensions for the $SU(2)$  spin chain from $\mathcal{N}=4$ SYM and rewrite them in terms of a spin chain for bosons that satisfy the Cuntz oscillator relations. The truncation to the $SU(2)$ spin chain follows straightforwardly from the work \cite{MZ}, which computed the $SO(6)$ spin chain Hamiltonian. The main observation needed for the calculation is that for holomorphic operators made of scalars,  the contributions from D-terms, the gluon exchange and the self energy corrections cancel against each other \cite{D'Hoker:1998tz} so that only F-terms can contribute to the anomalous dimension. The superpotential of $\mathcal{N}=4$ SYM, $W= \Tr(X[Y,Z])$, gives the F-terms the structure
\begin{equation}
\hbox{F-terms} = g_{YM} \Tr(F_X [Y, Z] + F_Y[Z,X]+ F_Z[X,Y])+ \text{c.c.}
\end{equation}
plus the kinetic term $\Tr(F^* F)$. We keep the factor of $g_{YM}$ to be able to count loops, but other numerical coefficients are dropped.  
Although the $F$ fields are auxiliary variables, it is convenient to keep them in the Feynman diagrams.  Now we let the anomalous dimension computation proceed on a word of the type 
\begin{equation}
\label{eq:closedspinchain}
\ket{n_1, n_2, n_3 \dots, n_k}= \Tr(Y Z^{n_1} Y Z^{n_2} YZ^{n_3} \dots YZ^{n_k})
\end{equation}
Notice how we choose the labeling in terms of the number of $Z$ in between the $Y$. The standard convention would be to name these states as  a spin chain with $SU(2)$ indices at each position, which are an up state $Y\simeq \ket{\uparrow}$ and a down state $Z\simeq \ket{ \downarrow}$ \cite{MZ}. Thus, the map between the two conventions for labeling states is given by
\begin{equation}
\ket{n_1, n_2, n_3, \dots} \simeq \ket{\uparrow, \downarrow^{\otimes n_1}, \uparrow, \downarrow^{\otimes n_2}, \uparrow, \downarrow^{\otimes n_3}, \dots}
\end{equation}
The cyclicity of the trace enforces periodic boundary conditions on the spin chain. Thus we will refer to operators of the the form \eqref{eq:closedspinchain} as closed spin chains. Open spin chains will be discussed later in this section.

A term of the form $F^*_X[\bar Z, \bar Y]$ can be contracted (in the planar approximation) with two consecutive letters that are different, $Z$ and $Y$ more precisely, in a spin chain and it replaces it with an $F_X^*$.  If we change the order in which  the two letters appear we get a minus sign. We then get an intermediate state with the $F^*_X$ label. In a Hamiltonian formalism $\bar{Z},~\bar{Y}$ can destroy $Z,~Y$ letters (by contractions), so they act as lowering operators, whereas $Z,~Y$ fields create $Z,~Y$ letters in our words (the list of operators). We can write a lowering operator for the $Z$ letters as $\partial_Z$. Similarly, we can write a lowering operator for the $Y$ letter as $\partial_Y$. In general these satisfy the Heisenberg algebra in the Bargmann representation.

\begin{figure}[ht]
\includegraphics[width=7 cm ]{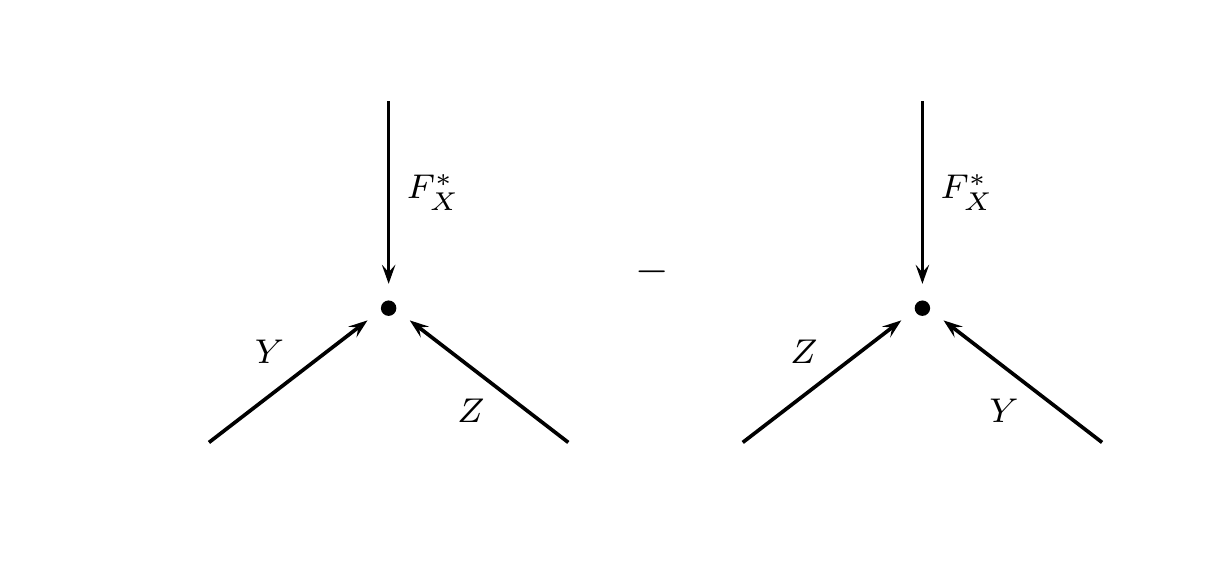}
\caption{Feynman diagrams with auxiliary fields in the final state}\label{fig:fd1}
\end{figure}

So we have that  $F^*_X[\bar Z, \bar Y]\simeq F_X^* [\partial_Z, \partial_Y]$.
Diagramatically this is given by figure \ref{fig:fd1}
where we see how the words $YZ$ and $ZY$ get converted from the initial state into an auxiliary field for the $X$ superfield.

\begin{figure}[ht]
\includegraphics[width=7 cm ]{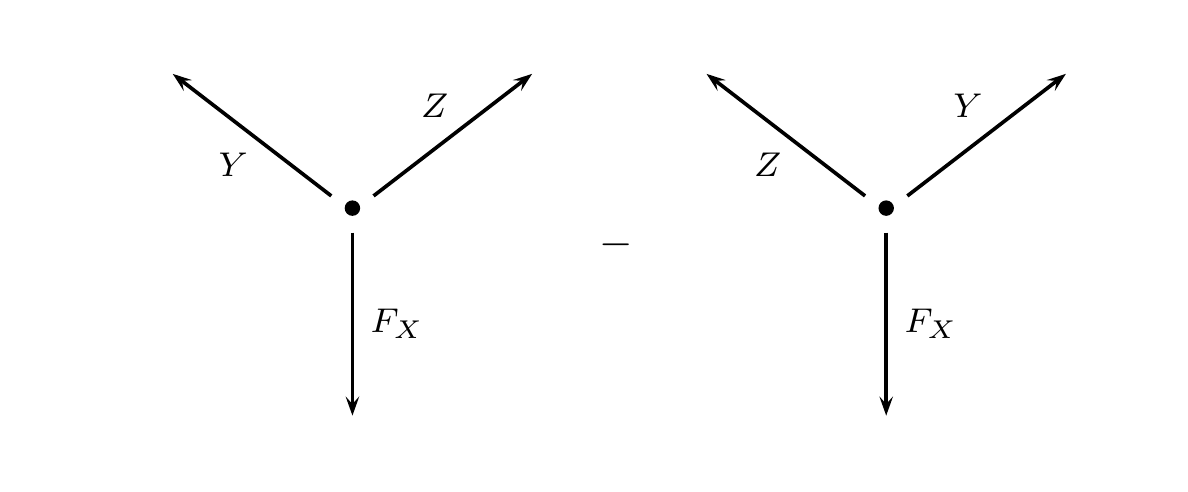}
\caption{Feynman diagrams with auxiliary fields in the initial state}\label{fig:fd2}
\end{figure}

When we integrate out the $F$ fields, the $F^*_X$ needs to be contracted with an $F_X$, as required by the kinetic term, but the term with $F_X$ in the action is exactly $F_X[Y,Z]$. These diagrams are given by
figure \ref{fig:fd2}

The end result is that the anomalous dimension of these operators to one loop order is generated by the term 
\begin{equation}
H_{\text{eff}} = g_{YM}^2 \mathop{:}\Tr( [Y,Z][\partial_Z, \partial_Y])\mathop{:}
\end{equation}
which is just the concatenation of the diagrams in the top with the ones in the bottom modulo factors of $2\pi$ and numerical constants that define the precise normalizations of contractions in the theory. The normal ordering indicates that the derivatives should not act on the fields inside the normal ordering. When we are working to one loop order the precise conventions for couplings do not matter as much as we are not comparing different orders in perturbation theory. This is the effective one loop Hamiltonian described in \cite{Beisert:2003tq}. The effective Hamiltonian operates on words as described above in the following form (so long as one does not have zero occupation numbers on some of the $n_i$)
\begin{align}
H_{\text{eff}} \ket{n_1, n_2, n_3 \dots, n_k} &= g_{YM}^2 N \sum_{i=1}^k 2 \ket{\dots, n_{i-1},  n_i, n_{i+1} \dots } \\
& - \ket{ \dots, n_{i-1}+1,  n_i-1 , n_{i+1} \dots } - \ket{ \dots, n_{i-1},  n_i-1 , n_{i+1}+1 \dots } \nonumber
\end{align}
The sum over the last and first term require us to use the identification that $n_{k+1}=n_1$ and $n_0 = n_k$. This makes the spin chain periodic. 

It is convenient to introduce a set of raising and lowering operators $a^\dag$ and $a$, such that $a^\dagger \ket{n} = \ket{n+1}$ for $n \geq 0$, $a \ket{n} = \ket{n - 1}$ for $n > 0$, and $a\ket{0} = 0$ so that negative occupation numbers are not allowed. Such a set of creation and annihilation operators satisfy the Cuntz algebra\footnote{The Cuntz algebra is related to a deformation of the canonical harmonic oscillator algebra.}, that is, $a a^\dagger = I$, and $a^\dagger a = I - P_{0}$, where $I$ is the identity and $P_{0} = \ket{0}\bra{0}$ is the projector onto the zero occupation state. We can define a similar set of operators that act on each individual site of the spin chains via $a_i = I^{\otimes^{i - 1}} \otimes a \otimes I^{\otimes^{k - i}}$ and likewise for $a^\dag_i$, which indeed commute with each other. These satisfy the commutation relations $[a_i, a_j^\dag] = \delta_{ij} I^{\otimes^{i - 1}} \otimes P_0 \otimes I^{\otimes^{k - i}}$. Using this representation, the effective Hamiltonian can be written in the following convenient form
\begin{equation}
H_{\text{eff}} = g_{YM}^2 N\sum_{i = 1}^k  2 a_i^\dagger a_i - a_{i-1}^\dagger a_i - a_{i+1}^\dagger a_i 
\end{equation}
using periodic identifications of the sites. The first term can be thought of as the energy for staying in place, whereas the other two terms can be interpreted as particles hopping out of site $i$ into site $i+1$ or $i-1$. This Hamiltonian can also be written as follows
\begin{equation}
\label{eq:cuntzchain}
H_{\text{eff}} = g_{YM}^2 N\sum_{i = 1}^k  (a_{i+1}^\dagger - a_i^\dagger)(a_{i+1}- a_i)
\end{equation}
which shows that it is a sum of squares. 

Notice that because $a a^\dagger = I$, all operators should naturally be written as linear combinations of objects in normal ordered form $\hat S_{kn} = (a^\dagger)^k a^n$. 
It is easy to show that $\hat S_{nn}= (a^\dagger)^n a^n = 1 - \sum_{m=0}^{n - 1} P_m$, where the $P_m$ are the projectors on the state with occupation number $m$.
It is clear that $\hat S_{nn}\ket{m} = 0$ if $n > m$, and that otherwise $\hat S_{nn} \ket m = \ket m$. Thus the occupation number is given by the following expression
\begin{equation}
\hat N = \sum_{n=1}^\infty \hat S_{nn} =\hat N^\dagger
\end{equation}
It is easy to show that $\hat N_{\text{tot}} = \sum_i \hat N_i $ commutes with the Hamiltonian. This follows straightforwardly from the commutation relations $[\hat N, a^\dagger]= a^\dagger$, which are themselves straightforward to prove.  

There is one extra ingredient that we needed: the cyclic property of the trace in the setup. This condition will not be necessary when we talk about open spin chains, but we need it in the case of closed spin chains. The condition of cyclicity of the trace is related to the level matching constraints in string theory \cite{BMN}. This condition is implemented in the allowed wave functions. Notice that rotations in the trace need to take $Y$ letters to $Y$ letters in order to match the labeling of the states that we have,  so the set of cyclic permutations just permute cyclically the occupation numbers $n_i\to n_{i+1}$. This operation commutes with the Hamiltonian. 
We require that the constraint be satisfied in the following way
\begin{equation}
\psi(n_1, n_2, \dots, n_k) = \braket{n_1, \dots n_k} \psi = \psi( n_2, n_3, \dots, n_k, n_1)
\end{equation}

It is easy to show that there are a huge number of vacuum states with zero value for the one loop energy. All we need to do is to find states for which $a_{i+1}= a_i$ for all $i$. These are operator equations, but we can turn these into c-number equations. The idea is to introduce the notion of a coherent state for the site $i$, by declaring that 
\begin{equation}
a_i \ket{z_i} = z_i \ket{z_i}, \label{eq:chostat}
\end{equation}
so that the corresponding state is an eigenstate of the lowering operators with eigenvalue $z_i$. Then solving $a_{i+1} = a_i$ boils down to setting $z_{i+1}=z_i$.

For the state $\ket{z}$, we can easily solve the coherent state equations to find that 
\begin{equation}
\ket{z} = \sum_{k=0}^\infty z^k \ket{k}
\end{equation}
The state is normalizable if
\begin{equation}
\braket{z}{z} = \sum_{k=0}^\infty |\bar{z}z|^k < \infty 
\end{equation}
The sum converges if $|z|<1$, and in that case it converges to $\braket{z}{z} = (1 - \bar{z}z)^{-1}$. The vacua are therefore characterized by the states where $z_{i+1} = z_i = z$ for all $i$, and these are given by
\begin{equation}
\ket{\psi} = \ket{z, z, \dots z}
\end{equation}
It is easy to check that all of these satisfy the cyclicity property of the trace. 
Notice that these are valid states for all $z$ such that $|z| < 1$. Notice also that these states with zero energy do not have a fixed occupation number, but that if we project these states to a fixed occupation number $N_{\text{tot}}$ (which remember, commutes with the one loop Hamiltonian) 
then we find that the wave function is such that $\psi(n_1, n_2, \dots, n_k) = 1$ for all $n_i$ with $\sum n_i = N_{\text{total}}$, and that there is exactly one such state. This matches with the chiral ring computation of the ${\cal N} = 4$ SYM, where there is a unique single trace element of the schematic form $\Tr(Z^{N_{\text{tot}}} Y^k)$.  

Now, we will proceed with the computation of the Cuntz chain Hamiltonian for the case of an open string attached to a pair of giant gravitons. The formulation found in \cite{Berenstein:2013md} shows that for a single giant graviton, it is natural to use the basis of operators given by
\begin{equation}
\det(Z-\lambda) \Tr\left( \frac{1}{Z-\lambda} Y Z^{n_1} Y \dots Z^{n_k} Y\right)
\end{equation}
which would represent a single string starting and ending on the same giant gravitons. We can call this state by the label $\ket{\lambda; n_1, \dots n_k}$.  Notice that the trace now has a preferred site: the one where $\lambda$ shows up. Thus we do not impose the cyclic property on the wave functions for the $n_i$. 

When dealing with multiple giant gravitons, we want the start and the end of a string to be in different giant gravitons, so that the boundary condition on the left can take a different value than the boundary condition on the right. Just as in \cite{Berenstein:2013md}, it is simpler to work in the supersymmetric $\BZ_2$ orbifold of ${\cal N}=4$ SYM, rather than in ${\cal N}= 4 $ SYM directly. This corresponds to a $U(N)\times U(N)$ quiver theory with ${\cal N}=2$ SUSY in four dimensions. The chiral superpartners of the vector fields will be called $Z,~\tilde{Z}$, while the matter hypermultiplets between the two gauge groups will be made of $X,~Y$ chiral
fields. 
The corresponding state will be given by
\begin{equation}
\label{eq:openspinchainorbifold}
\det(Z-\lambda) \det(\tilde Z-\tilde \lambda)\Tr\left( \frac{1}{Z-\lambda} Y_{12} \tilde{Z}^{n_1} Y_{21} Z^{n_2} Y_{12} \tilde{Z}^{n_3} \dots Z^{n_k} Y_{12} \frac 1{\tilde{Z} - \tilde{\lambda}} X_{21}\right)
\end{equation}
Where we note that we need another string to go back to the original giant graviton, which we have made out of a single $X$. The labels $Y_{12}$ indicate that the $Y$ is a bifundamental in the $(N_1, \bar N_2)$ representation of the $U(N_1)\times U(N_2)$ orbifold group (with $N_1=N_2=N$ numerically), whereas the $Y_{21}$ is in the $(\bar N_1, N_2)$ representation.

The idea of using a different letter for the string that heads back is that we can isolate the 
contributions to the anomalous dimension from the $Y,~Z$ interactions as described above and forget the ones that come from the $Z,~X$ interactions. That way there is no double counting of contributions. To get the full result, we add another copy of the computation for the $X$ word. Notice also that in the spin chain the $Z$ and the $\tilde{Z}$ alternate between each other, so we are forced to have an even number of sites. This does not affect the boundary conditions on the spin chain that we want to derive. We will label the state described above by the convention $\ket{\lambda, \tilde \lambda; n_1, \dots n_k}$, so we are making it clear that the string goes between a brane located at $\lambda$ and another brane located at $\tilde{\lambda}$ and that we are really only considering the string made of $Y$ fields for our problem. 

There are various results that need to be put together to calculate the boundary contributions to the anomalous dimension of the above state. We need to compute the following objects:
\begin{equation}
\bra{\lambda, \tilde \lambda; n'_1, \dots n'_k} H_{\text{eff}} \ket{\lambda, \tilde \lambda; n_1, \dots n_k}\label{eq:Heffmel}
\end{equation}
where the states are normalized to have unit norm.
The first question, is therefore to compute the norm of the bare states above, to leading order in a $1/N$ expansion. 

This answer is given by 
\begin{equation}
|\ket{\lambda, \tilde \lambda; n_1, \dots n_k}|^2= N^{k+\sum n_i} (N-\lambda\lambda^*)(N-\tilde \lambda \tilde \lambda^*)   \exp( \lambda \lambda^*+\tilde\lambda\tilde\lambda^*) \label{eq:norm}
\end{equation}
and result from combining the results of \cite{Berenstein:2013md} with planar contractions of the words of the spin chain. The planar contractions of $W^{a}_{b} = (Y\tilde{Z}^{n_1}YZ^{n_2}Y\tilde{Z}^{n_3}\cdots Z^{n_k}Y)^{a}_{b}$ and its complex conjugate $\bar W^{\tilde b}_{\tilde a}$ in the leading planar approximation give a result proportional to $\delta^a_{\tilde a}\delta^{\tilde b}_b$. 
The factor of $N^{k+\sum n_i}$ is this proportionality factor. This counts the number of matrix contractions necessary to make the word $W$. The result is as if the composite word $W$ was acting as a single $Y$, but with a different normalization factor. That other contributions are subleading in powers of $1/N$ was shown in \cite{Berenstein:2003ah}. The reason these factors of $N$ are important is that the effective Hamiltonian changes the number of the $n_i$ at the edges, namely $n_1, n_k$ in such a way that they can exit the region sandwiched between the $Y$, thus the results in \eqref{eq:Heffmel} can have different powers of $N$ and this affects the naive planar counting. 

To compute the matrix elements of $H_{\text{eff}}$ using un-normalized states we therefore need to divide by the norm of the states carefully. For example, if one has in an un-normalized basis that 
\begin{equation}
H_{\text{eff}} \ket a= \sum_b \tilde H_{ba} \ket b
\end{equation}
Then we have that in a normalized basis
\begin{equation}
H_{\text{eff}} \frac{\ket a}{|\ket a|} = \sum_b \frac{|\ket b|}{|\ket a|}\tilde H_{ba}\frac{\ket b}{|\ket b|} 
\end{equation}
So 
\begin{equation}
H_{ba} = \frac{|\ket b|}{|\ket a|}\tilde H_{ba}
\end{equation}

Considering that the effective Hamiltonian is described in \cite{Berenstein:2013md} by the following expression
\begin{equation}
H_{\text{eff}}= \mathop{:}\Tr[\bf{Z},\bf{Y}][\partial_{\bf Y},\partial_{\bf Z}]\mathop{:}
\end{equation}
where
\begin{eqnarray}
{\bf Z} = \begin{pmatrix} Z & 0\\
0&\tilde Z
\end{pmatrix}, & \quad & {\bf Y}=\begin{pmatrix} 0&Y_{12}\\
Y_{21}& 0
\end{pmatrix}\\
\partial_{\bf Z}=\begin{pmatrix}\partial_ Z & 0\\
0 & \partial_{\tilde Z}
\end{pmatrix}, & \quad & \partial_{\bf Y} = \begin{pmatrix} 0 & \partial_{Y_{21}}\\
\partial_{Y_{12}} & 0
\end{pmatrix}
\end{eqnarray}
we can pinpoint various contributions to the anomalous dimension. The first one, is where we take a $\tilde{Z}$ from $\tilde{Z}^{n_1}$ and move is to the left of the $Y_{12}$. We call that a hop-out interaction (following the conventions of \cite{deMelloetal}). The planar contribution to that term is captured by
\begin{equation}
-\mathop{:}\Tr(Z Y_{12} \partial_{\tilde{Z}} \partial_{Y_{12}})\mathop{:}
\end{equation}
Acting on the initial state gives a power of $N$ from contractions between the derivatives and the word $Y_{12} \tilde{Z}$.  Notice that the words go in opposite order than the way derivatives act on them (this is illustrated in \cite{Berenstein:2004ys}, particularly the section on matrix models). The state we get after this operation is given by
\begin{equation}
\det(Z - \lambda) \det(\tilde{Z} - \tilde{\lambda})\Tr\left(\frac{Z}{Z-\lambda} Y_{12} \tilde Z^{n_1-1} Y_{21} Z^{n_2} Y_{12} \tilde Z^{n_3} \dots Z^{n_k} Y_{12} \frac{1}{\tilde{Z} - \tilde{\lambda}} X_{21}\right)
\end{equation}
Now, in the term with the $Z$ pole in the trace we use the substitution $Z= (Z-\lambda)+\lambda$, generating two terms. One of them is $\lambda \ket{\lambda, \tilde \lambda; n_1-1, n_2, \dots, n_k}$, and the other one 
which has no pole anymore at $Z=\lambda$, which is to be considered as a single string state starting from brane two and ending on brane two. Such term counts as changing the number of strings and it is non-planar (this can also be checked by computing norms). The other term counts as planar, but proportional to $\lambda$. Thus, the end result is proportional to $N\lambda$. However, when normalizing the states, we see that the norm of the states changes as described in equation \eqref{eq:norm}, so that the result for normalized states with unit norm is actually given by following the recipe in equation \eqref{eq:Heffmel}, which involves the ratio of the norms of the states. The result is that for each $Z$ we create we attach a factor of $\sqrt{N}$ and for each $Z$ we annihilate we attach a $1 / \sqrt{N}$. When translating to the Cuntz oscillator basis, we can use $(\sqrt{N}a_i^\dag)$ and $(a_i / \sqrt{N})$ without having to recompute the normalizations of the states; using these replacements instead of $a_i^\dag$ and $a_i$ takes care of it for us. In this process we have one less $Z$ and so the hop-out interaction from the first element of the spin chain is given by the following extra contribution 
\begin{equation}
H_{\text{hop-out, left}}\simeq - g_{YM}^2 N \frac{\lambda}{\sqrt N} a_1
\end{equation}
Hermiticity ensures that the hop-in interaction is the adjoint of this operation, so we have that 
\begin{equation}
H_{\text{hop-in, left}}\simeq - g_{YM}^2 N \frac{\lambda^*}{\sqrt N} a^\dagger_1
\end{equation} 
Finally, there is one extra contribution to the left from acting with the term $\Tr(Z Y\partial_Y\partial_Z)$ where the derivative with respect to $Z$ acts on the giant graviton. Such terms are identical to those that were already computed in \cite{Berenstein:2013md}, and these are given by
\begin{equation}
g_{YM}^2 \lambda\lambda^*
\end{equation}
Such terms were called `kissing interactions' in \cite{deMelloetal}.

Putting it all together, we find that the open spin chain Hamiltonian on the left side of the spin chain is given by
\begin{equation}
\label{eq:sc}
H_{\text{eff}} \simeq g_{YM}^2 N \left[\left(\frac{\lambda}{\sqrt N} - a_1^\dagger\right)\left(\frac{\lambda^*}{\sqrt N} - a_1\right) 
+ (a^\dagger_1-a^\dagger_2)(a_1-a_2) + \dots \right]
\end{equation}
A similar term shows up in the right hand side, with $\lambda \to \tilde \lambda$ and $a_1\to a_k$. Notice that this is a simple generalization of equation \eqref{eq:cuntzchain} at the boundaries. This is a nearest neighbor interaction with hopping in and out of the chain at the boundaries. It is important to notice that since the parameter $\lambda$ is complex, there are phases associated to hopping in and out at the boundary. This is a simple generalization of the spin chain Hamiltonian found in \cite{Berenstein:2005fa,Berenstein:2007zf}. Notice that the Hamiltonian can be made to be the same as the one presented in that work if we choose $\lambda = \tilde \lambda = -\sqrt{N(1 - p / N)}$ in the notation of \cite{Berenstein:2005fa}. Notice that this result ends up having the same information content as the one found in \cite{deMelloetal} (particularly equations 3.7 and 3.8). All we have to do is interpret the parameter $\lambda$ in our expression in terms of raising and lowering operators associated to the momentum of the giant graviton. Since $\lambda$ is a coherent state parameter for a (inverted) harmonic oscillator, as shown in \cite{Berenstein:2013md}, we can think of $\lambda \simeq b$ and $\lambda^*\simeq b^\dagger$, for a harmonic oscillator pair. In this case, acting with a lowering operator actually increases the R-charge of the giant, and acting with the raising operator lowers the charge. We also have to be mindful of conventions with respect to signs. When we chose the operators $\det(Z-\lambda)$ as our giant graviton representatives, we get minus signs in the expansion in terms of subdeterminants. Those minus signs appear in the relative sign between $\lambda^*$ and $a_1$ in the expressions above. If we would have chosen the operators $\det(Z+\lambda)$ instead, we would have gotten the result above with various signs changed. Those sign differences would reproduce the results of \cite{deMelloetal} exactly, while changing from Cuntz oscillators to ordinary oscillators would account for the numerical factors in the square roots appearing in equation (3.7), as well as the equation on page 23 describing the boundary Hamiltonian.

\section{Ground State for Open Spin Chain and Geometric Interpretation}
\label{sec:coh}

Our purpose in this section is to find the ground state for the Hamiltonian computed in equation \eqref{eq:sc}. 
\begin{equation}
H_{\text{spin chain}} \simeq g_{YM}^2 N \left[\left(\frac{\lambda}{
\sqrt N}-a_1^\dagger\right) \left(\frac{\lambda^*}{
\sqrt N}-a_1\right)+ (a^\dagger_1-a^\dagger_2)(a_1-a_2)+\dots \right]
\end{equation}
The idea is to use a trial wave function which is made of coherent states for the Cuntz oscillators as described in equation \eqref{eq:chostat} and to show that after minimizing with respect to the coherent state parameters that it is an eigenstate of the Hamiltonian. Thus, we use 
a label for the state as $\ket{z_1, \dots z_k}$, where the $z_i$ indicate coherent states for each Cuntz oscillator.
Using $a_i \ket {z_i} = z_i\ket{z_i}$ we find that when we evaluate the Hamiltonian
\begin{equation}
\bra{z_1, \dots z_k} H_{\text{spin chain}}\ket{z_1, \dots z_k}= g_{YM}^2 N \left[\left|\frac{\lambda^*}{
\sqrt N}-z_1\right|^2 + \sum_{i = 1}^{k-1} |z_i -z_{i+1}|^2+ \left|z_k - \frac{\tilde{\lambda}^*}{
\sqrt N}\right|^2\right]
\end{equation} 
which is a simple quadratic function of the $z_i$. When we minimize with respect to the $z_i$ parameters we find that
\begin{equation}
\frac{\lambda^*}{
\sqrt N}-z_1 = z_1 - z_2 = \dots = z_i -z_{i+1} = \dots = z_k-\frac{\tilde \lambda^*}{
\sqrt N}\label{eq:equal}
\end{equation}
Adding these together we find that
\begin{equation}
\frac{\lambda^*}{
\sqrt N}-\frac{\tilde \lambda^*}{
\sqrt N}= (k+1) (z_i -z_{i+1})
\end{equation}
so that 
\begin{equation}
z_i -z_{i+1}= \frac 1{k+1}\left(\frac{\lambda^*}{
\sqrt N}-\frac{\tilde \lambda^*}{
\sqrt N}\right)
\end{equation}
and the energy of this state is
\begin{equation}
E^{(1)}_0 = \frac{g_{YM}^2N}{k+1}\left|\frac{\lambda}{
\sqrt N}-\frac{\tilde \lambda}{
\sqrt N}\right|^2
\end{equation}
Here the superscript indicates that we are doing a one loop computation, and the subscript indicates the ground state energy.

There are various important things to notice. First, when $k=0$ (a chain with no sites), we reproduce the energy of the configurations calculated in \cite{Berenstein:2013md}, to show that we have consistency with the previous evaluation using the collective coordinate method. 
Also, if $\lambda=\tilde \lambda$, we get a state with zero energy, and we reproduce the results first deduced in \cite{Berenstein:2005fa}, with the same ground state. More importantly, consider the following observation. The following identity is an operator equation
\begin{equation}
\frac{\lambda^*}{
\sqrt N}-a_1+\sum ( a_{i}-a_{i+1}) + a_k-  \frac{\tilde \lambda^*}{
\sqrt N}= \frac{\lambda^*}{
\sqrt N}-\frac{\tilde \lambda^*}{
\sqrt N}
\end{equation}
which can also be applied to any state $\ket\psi$. If we let $\ket\psi_i = ( a_{i}-a_{i+1}) \ket\psi$, with 
$\ket\psi_0$ and $\ket\psi_{k}$ defined to contain the other two summands in the operator equation, we have the equality
\begin{equation}
\sum_{i=0}^{k} \ket\psi_i = \left( \frac{\lambda^*}{
\sqrt N}-\frac{\tilde \lambda^*}{
\sqrt N}\right)  \ket \psi
\end{equation}

Using the inequality between the quadratic mean and the arithmetic mean (suitably generalized to complex vector spaces), we find that
\begin{equation}
\frac 1{k+1} \sum_{i=0}^{k} |\ket\psi_i|^2 \geq \left|\frac 1{k+1} \sum_{i=0}^k \ket\psi_i\right|^2 = \left| \frac{\lambda}{
\sqrt N}-\frac{\tilde \lambda}{
\sqrt N}\right|^2 \frac1{(k+1)^2}\braket\psi\psi
\end{equation}
This follows from convexity of the function $|\ket\psi|^2$ on a complex vector space. 
This inequality is saturated only if the $\ket\psi_i$ are all the same vector. 
For a normalized ket such that $\braket\psi\psi=1$, we can translate this into the following inequality
\begin{equation}
\bra \psi H_{\text{spin chain}} \ket\psi = g_{YM}^2 N \sum_{i=0}^{k} |\ket\psi_i|^2 \geq  \frac{g_{YM}^2 N}{(k+1)}\left| \frac{\lambda}{
\sqrt N}-\frac{\tilde{\lambda}}{
\sqrt N}\right|^2 = E^{(1)}_0 \label{eq:ineq}
\end{equation}
which shows that the energy for any other state is higher than the one for the coherent state we found. This shows we have in principle found the ground state for the system. The only thing we are still left to show is that the state has finite norm, this is, the $z_i$ are such that $|z_i|<1$. This is also easy to show. After all, the $\lambda$ are required to have norm less than $\sqrt N$ \cite{Berenstein:2013md}. Also, the equation \eqref{eq:equal} shows that the $z_i$ are equidistant of each other, and they form an array of $k$ evenly spaced points stretching between $\lambda^* N^{-1/2}$ and $\tilde \lambda^* N^{-1/2}$. Thus all of the $z_i$ are in the unit disk and the state is normalizable.

\begin{figure}[ht]
\includegraphics[width=7 cm ]{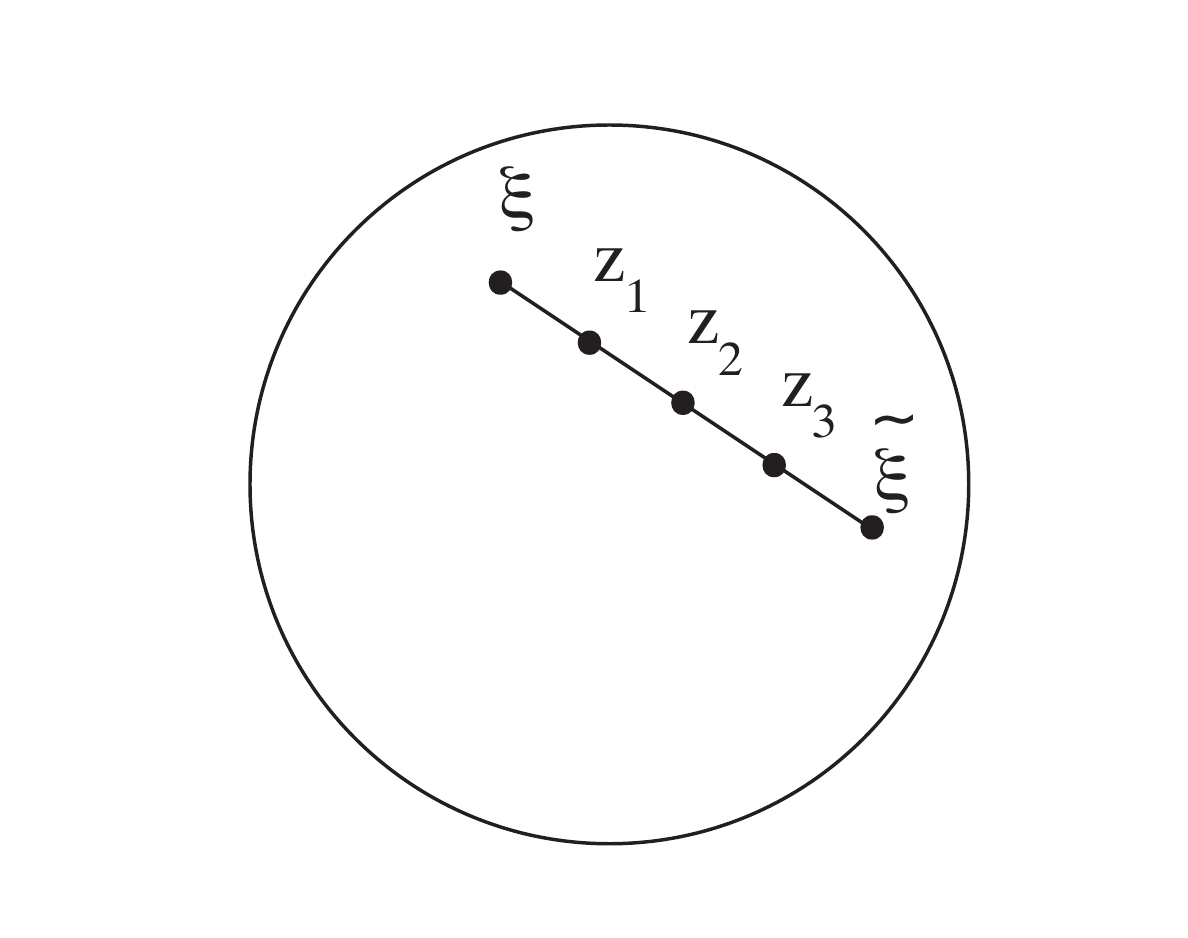}
\caption{Geometric layout of the $z_i$ in the ground state, as an interpolating chain of complex numbers between $\xi$ and $\tilde \xi$.}\label{fig:cgs}
\end{figure}

What we see is that the $z$ coordinates are very closely related to the $\lambda$ coordinates characterizing giant gravitons. It is convenient to introduce coordinates for the giant gravitons
$\xi= \lambda^* N^{-1/2}$ and $\tilde \xi=\tilde \lambda^* N^{-1/2}$. These coordinates are the complex conjugates of similar named variables in \cite{Berenstein:2013md}. This can be interpreted geometrically in the  figure \ref{fig:cgs}

The spin chain energy can be evaluated for a general state of the  coherent state basis and the energy can be interpreted as a collection of variables $z_i$ in the unit disk, with $z_0 = \xi$ and $z_{k+1} = \tilde \xi$, and the expectation value of the energy of such a general configuration is a sum of distances squared in the complex plane, namely
\begin{equation}
E (z_0, \dots , z_{k+1}) \simeq g_{YM}^2 N \sum_i |z_{i+1}-z_i|^2 \label{eq:scenergy}
\end{equation}
which is a geometric equation. Although $z_0$ and $z_{k+1}$ are really giant graviton coordinates, the energy of the state does not really treat them differently than the other $z_i$ (other than being at the endpoints). We can reverse the logic and state that  all the $z_i$ could be treated as if they are D-brane coordinates of some sort. In this sense, the formula for the mass squared starts looking like a sum over contributions where impurities are stretched between successive D-branes, with gauge invariance requiring that for each incoming string to a brane there is an outgoing string. This type of interpretation gives further evidence for the idea that geometry at strong coupling can be understood 
in terms of open strings stretching between gases of eigenvalues (D-branes) as espoused originally in \cite{Berenstein:2005aa}, and also makes contact with the calculations in \cite{Berenstein:2005jq, Hatsuda:2006ty} which required a saddle point integral to obtain a geometric interpretation. It further improves on these by allowing specific D-brane endpoints. Indeed, we find for the ground state an exact calculation at one loop, rather than an approximation.

To make further contact with the geometric interpretation of giant gravitons described in \cite{Berenstein:2013md}, we would like to write an effective action for the spin chain, whose Hamiltonian is given by equation \eqref{eq:scenergy}. For this, we need to compute the Berry phase associated to the coherent state basis for the spin chain. This was done originally in \cite{Berenstein:2005fa}, but the result was written in a different coordinate system. For this, we need the following result for unnormalized Cuntz coherent states
\begin{equation}
\braket{\tilde z} z= \sum_{k=0}^\infty (\tilde z^* z)^k = \frac 1{1-\tilde z^* z} \label{eq:cohoverlap}
\end{equation}
Then, we can take the limit
\begin{equation}
\lim_{\tilde z\to z(t)} \frac 1 {\braket{\tilde z} {z(t)}} i\partial_t \braket{\tilde z} {z(t)} =  \frac{i \bar z\dot z}{(1-\bar z z)}
\end{equation}
This gives us a Berry connection on the set of states parametrized by $\ket z$. If we choose to work with normalized states, we use the following definition for the Berry phase
\begin{equation}
\lim_{\tilde z\to z(t)} \frac 1 {\sqrt {\braket{\tilde z} {\tilde z}}} \partial_t \left(\frac 1{\sqrt{\braket {z(t)} {z(t)}}} \braket{\tilde z} {z(t)}\right)= \frac i2 \frac{\bar z \dot z - \dot {\bar z} z}{(1-\bar z z)}
\end{equation}
which differs from the above by a total derivative. The Berry connection associated to the holomorphic variable $z$ gives rise to a metric for the $z$ coordinate, which turns out to be the metric of the Poincar\'e disc.

Now, apart from the one loop correction to the Hamiltonian, there is the free field theory result for the energy of the spin chain. This energy counts the number of $Z$ and $Y$ inside the chain. Counting $Y$'s is easy, and it is given by $k+1$. On the other hand, we can only count the $Z$ as an average. Indeed, this is easiest to do by noticing that equation \eqref{eq:cohoverlap} is a generating series for the amplitudes of the different $\ket n $
states in a coherent state parametrized by $z$.   
\begin{equation}
\langle N_z \rangle=  \lim_{\tilde z \to z} (\braket{\tilde z} z)^{-1} z \partial_z \braket{\tilde z} z = \frac{ \bar z z}{1-\bar z z}
\end{equation}
If we write the action for the coherent states in the standard representation
\begin{equation}
S_{\text{eff}}= \int dt\, \bra {z(t)} i \partial_t \ket {z(t)} - \int dt\, \langle H \rangle 
\end{equation}
we find that 
\begin{equation}
S_{\text{spin chain}} = \int dt \left[\frac{i}{2}\sum_{i=1}^k \frac{\bar{z_i} \dot{z_i} - \dot{\bar{z}}_i z_i}{(1-\bar z_i z_i)} - (k+1) - \sum_{i=1}^k \frac{ \bar z_i z_i}{1-\bar z_i z_i} 
-  A g_{YM}^2 N \sum_{i=0}^k |z_{i+1}-z_i|^2  \right] 
\end{equation}
where $A$ is the normalization constant for the one loop result which includes factors of $2\pi$ etc. The terms at the end include $z_0 = \xi,~z_{k+1} = \tilde{\xi}$. If we include the giant gravitons in the action (and the string stretching back from the $\tilde{\xi}$ giant to the $\xi$ giant) as obtained in \cite{Berenstein:2013md}, being careful with the fact that $\xi$ is now defined to be the complex conjugate of the variable $\xi$ defined in that paper, we find that up to one loop order the full effective action of the giant gravitons with a spin chain of strings attached is given by
\begin{align}
S_{\text{tot}} &= N \int dt \left[\frac{i}{2}(\xi\dot{\bar{\xi}} - \bar{\xi} \dot{\xi}) - ( 1-\bar \xi \xi) \right]+N \int dt \left[\frac i{2} (\tilde{\xi}\dot{\bar {\tilde{\xi}}} - \bar{\tilde{\xi}}\dot{\tilde{\xi}}) - (1 - \bar{\tilde\xi}\tilde{\xi}) \right]\\
&\qquad - \int dt \left( 1+ A g_{YM}^2 N |\xi -\tilde{\xi}|^2\right)+ S_{\text{spin chain}}
\end{align}
The term $1+ A g_{YM}^2 N |\xi -\tilde \xi|^2$ is from the return string which we have labeled with an $X$ word (it is a spin chain with zero sites).

For the time being, let us look at the equations of motion of the $z_i$ and $\xi,~\tilde{\xi}$ in the case $g_{YM}^2=0$. We easily find that
\begin{align}
\dot{\xi} = -i \xi, &\quad \dot{\tilde{\xi}} = - i \tilde{\xi} \\
\dot{z_i} =& -i z_i 
\end{align}
so that the collective coordinates $\xi,~\tilde{\xi},~z_i$ rotate at the same angular speed. This means that in the free field theory limit we have that $|z_i -z_{i+1}|^2$ is constant. When we minimize the one loop term with respect to the $z_i$, we find a correction to the equations of motion, but on the ground state of the spin chain this correction vanishes for the $z_i$. The only true correction to the equations of motion is the 
correction to the equation of motion of $\xi,~\tilde{\xi}$. This correction is suppressed by $1/N$, so it can be considered as back-reaction of the brane to the presence of strings attached to it. If we consider small fluctuations of the $z_i$ around their ground state, the fact that the $\xi$ motion can only be corrected to order $1/N$ means that we can treat it as a Dirichlet boundary condition, namely $z_0,~z_{k+1}$ have a fixed motion described by the free giant graviton solution to the equations of motion.

Notice also the following: although the coordinates $\xi,~\tilde{\xi},~z_i$ parametrize the same disk, and they appear on similar footing in the interacting Hamiltonian, so that they share the equations of motion in the free field limit, the associated phase space for the variables is different: the Berry connection for the variables $\xi,~\tilde{\xi}$ gives rise to a flat metric in the $\xi$ coordinate, while the  Berry phase in the $z$ coordinate lead to a Poincar\'e disk metric. This indicates that the $\xi$ and the $z$ are in the end representing different objects: D-branes and pieces of strings. 
Also notice that when we take the D-branes to the edge of the disk we recover exactly the one loop dispersion relation for giant magnons and their bound states made of $k+1$ defects \cite{Dorey:2006dq}. We also find that the wave function  characterized by the $z_i$ is exactly of the Bethe ansatz type at complexified momentum and where the S-matrix coefficients for various re-orderings of the momenta vanish. It is natural to identify $z\simeq \exp(ip)$ at complex values of $p$ to make this identification.

Now, we can try to apply this result to other setups. For example, consider the problem of open spring theory, where at the end of a long computation to one loop order in a basis of operators made of Schur polynomials \cite{Corley:2001zk} with string decorations \cite{Balasubramanian:2004nb}, one obtains an effective Hamiltonian which is similar to a discretized hopping on a lattice \cite{Koch:2010gp}. Diagonalizing this problem leads to an effective theory of free harmonic oscillators \cite{deMelloKoch:2011ci}. The springs are made of a string stretching between two giants with a single $Y$ (a spin chain with zero sites). The main advance of \cite{Berenstein:2013md} was to simplify these computations by introducing the collective coordinate approach. In this paper we see that for strings (spin chains) made with $k+1$ Y's stretching between the giants, where we get a spin chain with $k$ sites, the one loop computation of the spring energy is reduced by $1/(k+1)$ relative to the case with a single $Y$ defect. This is similar to taking $k+1$ springs and connecting them in a series configuration. To connect them in parallel one would use many strings stretching between the giants. Thus we see that these computations generate more configurations for the open spring theory program that lead to a system of harmonic oscillators in the one loop approximation of anomalous dimensions.

\section{The Cuntz Hamiltonian at Two Loops}
\label{sec:2-loop}

So far we have studied the one loop effective Hamiltonian for strings stretching between giant gravitons, and we have computed their ground state. 
We have seen that the energy of the state can be understood geometrically. It is interesting to consider the higher loop computation for various reasons. 
Consider for example the case of the higher loop computation of BMN states
\cite{Gross:2002su,Santambrogio:2002sb}. The goal there was to show that the higher loop order corrections matched the string theory result for energies \cite{BMN} and to show that a low energy effective field theory on the worldsheet would appear. One can follow the same rationale here, and ask what is the effective Hamiltonian to two loop order and try to understand how the Hamiltonian and its ground state are organized.

Notice that it is simple to understand the giant magnon dispersion relation as arising from saturating a BPS constraint for a centrally extended $SU(2|2)$ algebra \cite{Beisert:2005tm}, and this also applies for their bound states \cite{Dorey:2006dq}. In general setups, like strings in flat space, one notices that the central charge extension of supersymmetry vanishes for closed string states on trivial topologies (those that have no fundamental group). This is not generally so for open strings. Indeed, if one considers D-branes in a flat background, the position separation between D-branes is a central charge. This is because the separation vector between two D-branes  can be thought of as being T-dual to momentum \cite{Dai:1989ua}. Since momentum appears as a central charge for supersymmetry (this is simplest to see by dimensional reduction), in general the central charges of excitations can be physical in the open string sector even if they are confined for closed strings. Indeed, one can check that this central charge extension property applies to solitons as well as fundamental particles in field theory \cite{Witten:1978mh}, and understanding this is the basis for dualities in $\mathcal{N}=2$ theories in four dimensions \cite{Seiberg:1994rs}. Hence, the fact that for the closed (long) string excitations one finds that the spectrum of excitations is characterized by building blocks that carry a central charge, this suggests that the spectrum of open strings might actually measure this charge directly in the ground state of the open string. Looking at the results we found in the previous section, it suggests that the string central charge is computed exactly by the difference of the coordinates of the end D-branes in the droplet plane. This is because we found that our computation of the one loop Hamiltonian was a sum of squares, and we obtained an inequality for the spectrum of the Hamiltonian at the end in equation \eqref{eq:ineq}, which depended only on the number of sites (namely, $k+1$) and the vector separation between the branes $ \lambda -\tilde \lambda$. Obtaining additional evidence to identify $ \lambda -\tilde \lambda$ as a central charge is important: it would tell us that
we can really start thinking of the giant gravitons as collective objects in the field theory in a way that resembles the Coulomb branch of $\mathcal{N}=4$ SYM more precisely. This is, after all, the expected low energy theory on the worldvolume of the giant gravitons themselves when they coincide. It would also tell us that the ground state of these strings might be computed by a BPS formula and extrapolating to strong coupling might be possible not only in principle, but in practice as well.

Consider also the problem of understanding the origin of this central charge from the point of view of string motion in $AdS_5\times S^5$.  For giant magnons this was understood in \cite{Hofman:2006xt} where it became clear what was the relation between the central charges of the spin chain and the geometric formulation in gravity, because the giant magnon was identified as a particular geometric configuration. It is important that relativistic effects play a very important role in making the match between the sigma model and the magnon dispersion relation possible. 
The geometric picture found this way is very similar to the one that was argued for earlier in field theory \cite{Berenstein:2005jq,Hatsuda:2006ty} using a saddle point approximation on a background made of a gas of eigenvalues. Obtaining further evidence for this geometric interpretation in field theory, and more precisely, the interpretation of geometries as being approximately described by eigenvalues forming some sort of quantum gas \cite{Berenstein:2005aa} whose excitations are bits of strings stretching between the eigenvalues is worthwhile. Finding exactly how this intuition ends up working in detail might give us a better understanding of how higher dimensional geometry emerges in field theory.

Finally, consider the problem of integrability of the spin chain. For closed string states, this has been argued in \cite{MZ,Bena:2003wd}, and that perhaps a one parameter family of integrable systems interpolate between these \cite{Dolan:2003uh}. A review of this program can be found in \cite{Beisert:2010jr}. For open strings, one generally expects that  general boundary conditions of the spin chain will not be integrable, but that for special boundary conditions the integrability might be preserved. It has been suggested that maybe integrability is preserved for general giant gravitons \cite{Berenstein:2006qk}, but only in the special case of the maximal giant graviton is there enough evidence for this (see for example \cite{Hofman:2007xp}), mostly because this is the one case where one can argue that the system might be solvable with a Bethe ansatz. In the case that the general giant graviton boundary condition is not integrable, understanding the structure of the  Hamiltonian to higher loop orders will help us understand the detailed dynamics of the strings better, by comparing it directly to the effective action of a string in a particular geometry (with $\alpha'$ corrections), rather than to the very special properties that make the system fully solvable in an integrable setup. Also, even if the system with giant graviton boundary conditions is not integrable, one might still be able to find some exact eigenstates of the spin chain Hamiltonian.

If we are going to do perturbation theory up to two loops, then we must take into consideration the relative coefficients between the contributions to the Hamiltonian at each loop order.
Here we use the conventions of \cite{Beisert:2003tq} and write the full Hamiltonain as
\begin{equation}
H = \sum_{n = 0}^\infty H_n, \quad H_n = \left(\frac{g_{YM}^2}{16\pi^2}\right)^{n}D_n 
\end{equation}
where $H_n$ is the $n$-loop contribution to the Hamiltonian.
The contributions to the Hamiltonian of the $SU(2)$ subsector of the full ${\cal N} = 4$ SYM up to two loop order were computed in \cite{Beisert:2003tq} to be
\begin{align*}
D_0 &= \mathop{:}\Tr(Z\partial_Z)\mathop{:} + \mathop{:}\Tr(Y\partial_Y)\mathop{:} \\
D_1 &= -2\mathop{:}\Tr[Y,Z][\partial_Y, \partial_Z]\mathop{:} \\
D_2 &= -2\mathop{:}\Tr\left[[Y,Z],\partial_Z\right]\left[[\partial_Y,\partial_Z],Z\right]\mathop{:} - 2\mathop{:}\Tr\left[[Y,Z],\partial_Y\right]\left[[\partial_Y,\partial_Z],Y\right]\mathop{:} \\
&\quad - 2\mathop{:}\Tr\left[[Y,Z],T^A\right]\left[[\partial_Y, \partial_Z], T^A\right]\mathop{:}
\end{align*}
where the $T^A$ are the generators of $U(N)$.
The action of $H_0$ on a closed spin chain counts the number of $Z$'s and $Y$'s.
In the Cuntz oscillator language this is the sum of the number operator (coming from the $Z$'s) and the number of sites (coming from the $Y$'s).
For an open spin chain the boundary terms yield an inverted harmonic oscillator for the collective coordinates of the giant graviton, as was shown in \cite{Berenstein:2013md}.
The one loop Hamiltonian was discussed in section \ref{sec:cuntz}. Putting back the numerical factors one has
\begin{equation}
H_{1} = \frac{1}{2}\left(\frac{g_{YM}^2N}{4\pi^2}\right)\left[\sum_i (a_{i+1}^\dag - a_{i}^\dag)(a_{i+1} - a_i) + \text{boundary terms}\right]
\end{equation}
where $H_1$ should be compared to equation \eqref{eq:sc}.
The numerical coefficient out front has been written suggestively for later reference.
Indeed the one loop Hamiltonian describes the sum of self energy like terms and the energy it takes to hop a $Z$ to the left or right.

Our goal here is to compute the highest order contribution to the two loop Hamiltonian in the Cuntz oscillator formalism.
The full calculation is long and tedious and so here we only point out some qualitative features regarding it leaving further details to the appendix.
Since $H_2$ is proportional to $D_2$, we need only consider the result by computing $D_2$.
Further, we will drop the overall factor of $-2$ in $D_2$.
All factors of proportionality will be restored at the end.
As was done at one loop order, we will consider the action of $D_2$ on a closed spin chain, then an open spin.
This amounts to computing the action of the operator on neighboring sites of the spin chain and then the boundary.

Recall that terms in the Hamiltonian can be thought of as search and replace functions in a text editor.
The derivatives delete letters and then replace them with whatever letters they were sitting next to.
The derivatives make a highest order contribution only when the derivatives are sitting next to each other and delete letters in a spin chain only when they match in {\it reverse} order.
This is due the fission / fusion rules associated with the generators of $U(N)$ (see appendix \ref{app:conventions}).
The number of factors of $N$ we pick up in this case is equal to the number of derivatives minus one.
Thus we only need to consider terms where all three derivatives are sitting next to each other.
There are sixteen such terms in $D_2$ after taking advantage of the cyclicity of the trace.
Note that for the closed spin chain we do not need to use the orbifold trick to simplify the calculation.

First we look at those term with two $\partial_Z$'s, of which there are eight.
As an example consider the term
\begin{equation}
\mathop{:}\Tr(ZYZ\partial_Z\partial_{Y}\partial_Z)\mathop{:}
\end{equation}
On neighboring spin sites this has the action
\begin{equation}
\Tr(\cdots Z^{n_i}YZ^{n_{i+1}}\cdots)\rightarrow N^2 \Tr(\cdots Z^{n_i - 1 + 1}YZ^{n_{i+1} - 1 + 1}\cdots)
\end{equation}
It is important to keep track of which $Z$'s are being deleted and added, as these correspond directly to the annihilation and creation operators.
In the Cuntz language this operator is
\begin{equation}
N^2 a_i^\dag a_{i+1}^\dag a_{i + 1}a_i
\end{equation}
Since $a_i$ commutes with $a_j^\dag$ for $i\neq j$, we are allowed to write the operator in this normal ordered form. This is a term with no hopping.

Let's consider a hopping term of the form
\begin{equation}
\mathop{:}\Tr(ZZY\partial_Z\partial_{Z}\partial_Y)\mathop{:}
\end{equation}
This has the action of
\begin{equation}
\Tr(\cdots Z^{n_i}YZ^{n_{i+1}}\cdots)\rightarrow N^2 \Tr(\cdots Z^{n_i + 2}YZ^{n_{i+1} - 2}\cdots)
\end{equation}
and generates a term in the Hamiltonian of the form
\begin{equation}
N^2 a_i^\dag a_i^\dag a_{i+1}a_{i+1}
\end{equation}
Now two $Z$'s have hopped to the left.
At one loop order we could only hop one $Z$ to the left or right.
At two loop order we are allowed to hop one or two $Z$'s to the left or to the right.
In order to generate one hop with two $Z$ derivatives, you need to delete and add one to the same site.
Such a term that appears and does the trick is
\begin{equation}
\mathop{:}\Tr(ZYZ\partial_Y\partial_Z\partial_Z)\mathop{:} \rightarrow N^2 a_i^\dag a_{i+1}^\dag a_{i+1} a_{i+1}
\end{equation}
The terms contributing to the Hamiltonian coming from two $Z$ derivative terms are
\begin{equation}
\label{eq:twopartialz}
\sum_{i} (a_{i+1}^\dag - a_i^\dag)(a_{i+1}^\dag( -a_i) + (-a_i^\dag)(a_{i+1}))(a_{i+1} - a_i)
\end{equation}
We later see where other terms come from to complete the square in the middle, yielding a nice quartic.

Next we look at terms with two $\partial_Y$'s, of which there are also eight.
There are three combinations of derivatives that appear: $\partial_Z\partial_Y\partial_Y$, $\partial_Y\partial_Z\partial_Y$, and $\partial_Y\partial_Y\partial_Z$.
Two $Y$'s will only be next to each other in a word if the corresponding site has zero occupation number.
Thus the first and last combination of derivatives must come along with a projection onto the zero state, which for convenience will be written as the commutator $[a_i, a_i^\dag] = P_{0_i}$.
The second combination of derivatives requires the corresponding site to have exactly occupation number one.
In the Hamiltonian this enters as the tensor that takes you from the one state to the zero state, that is, $\ket{0_i}\bra{1_i}$.
We can write this in the operator language as $[a_i, a_i^\dag]a_i$.
As an example we have the term
\begin{equation}
\mathop{:}\Tr(ZYY\partial_Z\partial_Y\partial_Y)\mathop{:}
\end{equation}
which has the action
\begin{equation}
\Tr(\cdots Z^{n_{i-1}}YYZ^{n_{i+1}}\cdots) \rightarrow \Tr(\cdots Z^{n_{i-1} + 1}YYZ^{n_{i+1} - 1}\cdots)
\end{equation}
and translates to
\begin{equation}
N^2 a_{i-1}^\dag[a_i,a_i^\dag]a_{i+1}
\end{equation}
Unlike the terms with two $Z$ derivatives, here we are allowed to hop a $Z$ over an empty site.
Computing the remaining seven terms and summing them up with the correct signs gives
\begin{equation}
\label{eq:twopartialy}
\sum_{i} (a_{i+1}^\dag - a_i^\dag)[a_i, a_i](-a_i + a_{i-1}) + (-a_i^\dag + a_{i-1}^\dag)[a_i,a_{i}^\dag](a_{i+1} - a_i)
\end{equation}
All terms with two $Y$ derivatives will generate terms with one commutator.
At this point it would seem that the two loop Hamiltonian will not have a nice form.
This will be resolved when we consider the last term of $D_2$.

The third term of $D_2$ can be simplified to
\begin{equation}
\mathop{:}\Tr\left[[Y,Z],T^A\right]\left[[\partial_Y, \partial_Z], T^A\right]\mathop{:} = 2ND_1
\end{equation}
The operator $D_1$ contains a factor of $N$ and so the two loop Hamiltonian has an order $N^2$ contribution which is only quadratic in the Cuntz operators.
Now the two loop Hamiltonian seems a bit awkward having terms both quadratic and quartic in the Cuntz operators.
To mitigate the awkwardness and package everything nicely, we insert a copy of the identity of the form $a^\dag a+ [a,a^\dag]$ for two different sites.
We write
\begin{align}
&\hspace{-1cm} 2\sum_i (a_{i+1}^\dag - a_i^\dag)(a_{i+1} - a_i) \nonumber \\
&= \sum_i (a_{i+1}^\dag - a_i^\dag)\lp a_i^\dag a_i + [a_i, a^\dag_i] + a_{i+1}^\dag a_{i+1} + [a_{i+1}, a^\dag_{i+1}]\rp(a_{i+1} - a_i) \nonumber \\
&= \sum_i (a_{i+1}^\dag - a_i^\dag)\lp a_i^\dag a_i + a_{i+1}^\dag a_{i+1} \rp(a_{i+1} - a_i) \nonumber \\
\label{eq:d2result}
&\quad + \sum_i (a_{i+1}^\dag - a_i^\dag) [a_i, a_i^\dag](a_{i+1} - a_i) + (-a_{i}^\dag + a_{i-1}^\dag) [a_i, a_i^\dag](-a_{i} + a_{i-1})
\end{align}
We put together the first sum of equation \eqref{eq:d2result} with \eqref{eq:twopartialz} and the second sum of equation \eqref{eq:d2result} with \eqref{eq:twopartialy} to obtain the nice form
\begin{align}
H_2 \propto N^2\sum_{i=1}^{k}(a_{i+1}^\dag - a_i^\dag)^2(a_{i+1} - a_i)^2 + (a^\dag_{i+1} - 2a_i^\dag + a_{i-1}^\dag)[a_{i},a^\dag_{i}](a_{i+1} - 2a_i + a_{i-1})
\end{align}
where the site labels have the identification $i + k \equiv i$.

At this point we have computed the two loop correction to the Hamiltonian for the closed spin chain.
Now we move on to the open spin chain by computing the boundary terms.
The orbifold trick does simplify things and so we will make use the operator \eqref{eq:openspinchainorbifold} for our open spin chain.
By symmetry we may focus solely on the left boundary of the open spin chain, that is, the graviton with collective coordinate $\lambda$ and the first site.
Thus for explaining this calculation we will take our open spin chain to have the schematic form
\begin{equation}
\label{eq:openspinchainschematic}
\det\lp Z - \lambda \rp\Tr\lp \frac{1}{Z - \lambda}Y_{12}\tilde{Z}^{n_1}WX_{21} \rp
\end{equation}
where $W$ is a word representing the rest of the spin chain.

The boundary terms coming from the term in $D_2$ proportional to $D_1$ have already been computed, but there are subtleties that will be discussed later.
The boundary of the spin chain will make a contribution only if one of the $Y$ derivatives in $D_2$ acts on the first $Y$ in the spin chain.
The $Z$ derivatives must act on the determinant,  or the first site.
However, if the $Z$ derivatives act on a site, then a boundary contribution is made only when we hop out some $Z$'s.
Our first example will show how these choices of where the  $Z$ derivative acts complicates things, but the orbifold trick makes things  simpler.

Consider the action of the following term from $D_2$ on \eqref{eq:openspinchainschematic}
\begin{equation}
\mathop{:}\Tr\lp Y_{12}\tilde{Z}\tilde{Z}\partial_{\tilde{Z}}\partial_{Y_{12}}\partial_{Z} \rp\mathop{:}
\end{equation}
Had we not used the orbifold trick, the $\partial_{\tilde{Z}}$ would be a $\partial_Z$ and have the option of acting on the determinant or the $(Z - \lambda)^{-1}$ inside the trace.
These terms would have split the spin chain and we would have thrown them away, but with the orbifold trick the bookkeeping becomes simpler.
Indeed the result is
\begin{align}
&\hspace{-1cm}\det\lp Z - \lambda \rp\Tr\lp \frac{1}{Z - \lambda}Y_{12}\tilde{Z}^{n_1}WX_{21} \rp \rightarrow \nonumber \\
& N\det\lp Z - \lambda \rp\Tr\lp \frac{1}{Z - \lambda}\frac{1}{Z - \lambda}Y_{12}\tilde{Z}^{n_1-1+1+1}WX_{21} \rp \\
&\quad - N\det\lp Z - \lambda \rp\Tr\lp \frac{1}{Z - \lambda} \rp\Tr\lp \frac{1}{Z - \lambda}Y_{12}\tilde{Z}^{n_1-1+1+1}WX_{21} \rp 
\end{align}
where in the first term $\partial_Z$ acted on the determinant and in the second term $\partial_Z$ acted inside the trace.
We already have made use of the equality $Z = Z - \lambda + \lambda$ and have thrown away terms of lower order.
Using the results in the appendix these terms collect into a nice derivative of $\lambda$:
\begin{align}
&\hspace{-3cm}\det\lp Z - \lambda \rp\Tr\lp \frac{1}{Z - \lambda}Y_{12}\tilde{Z}^{n_1}WX_{21} \rp \rightarrow \nonumber \\
&N\partial_\lambda \lb \det\lp Z - \lambda \rp \Tr\lp \frac{1}{Z - \lambda}Y_{12}\tilde{Z}^{n_1+1+1-1}WX_{21} \rp\rb
\end{align}
It was explained in \cite{Berenstein:2013md} how double trace terms will form derivatives, but here we have shown things explicitly to emphasize the care needed to get to the end result.
To translate this into the Cuntz language, we note that derivatives of $\lambda$ yield factors of $\lambda^*$ and we keep track of the normalizations using the results of section \ref{sec:cuntz}.
The contribution to the two loop Hamiltonian reads
\begin{equation}
N^2\lp \frac{\lambda^*}{\sqrt{N}}\rp a_1^\dag a_1^\dag a_1
\end{equation}
Just as double pole terms generate a single derivative with respect to $\lambda$, we can have triple pole terms and these will generate two derivatives with respect to $\lambda$.
Such a term will appear later on.
There are a total of eight terms with two $Z$ derivatives and all three derivatives sitting next to each that contribute to the boundary Hamiltonian.
The result of adding these together is
\begin{equation}
N^2\lp a_1^\dag - \frac{\lambda}{\sqrt{N}}\rp\lb\lp -\frac{\lambda}{\sqrt{N}}\rp a_1 + a_1^\dag\lp-\frac{\bar{\lambda}}{\sqrt{N}}\rp\rb\lp a_1 - \frac{\bar{\lambda}}{\sqrt{N}}\rp
\end{equation}

Now we deduce which terms with two $\partial_Y$'s can make order $N^2$ contributions.
A term containing $\partial_Y\partial_Z\partial_Y$ can only make a contribution if the $\partial_Z$ acts on the first site.
Otherwise a factor of $(Z - \lambda)^{-1}$ gets inserted and splits the spin chain.
In order to get a contribution from the boundary, we have to hop a $Z$ out of the chain.
There is only one such term in $D_2$.
\begin{equation}
\label{eq:firstboundarytwoy}
-\mathop{:}\Tr(Z_1Y_{12}Y_{21}\partial_{Y_{21}}\partial_{Z_2}\partial_{Y_{12}})\mathop{:}
\end{equation}
We can pick up at most one factor of $N$ from the $\partial_Z$ in terms with two $\partial_Y$'s.
To get the other factor of $N$, the $\partial_Y$'s must sit next to each other.
Although there are four such terms, one of them splits the spin chain.
The three remaining contributing terms are 
\begin{equation}
\label{eq:secondboundarytwoy}
\mathop{:}\Tr\lp Y_{12}Y_{21}Z_1\partial_{Y_{21}}\partial_{Y_{12}}\partial_{Z_1} - Y_{12}Z_2Y_{21}\partial_{Y_{21}}\partial_{Y_{12}}\partial_{Z_1} + Z_1Y_{12}Y_{21}\partial_{Z_1}\partial_{Y_{21}}\partial_{Y_{12}}\rp\mathop{:}
\end{equation}
Note that equation \eqref{eq:firstboundarytwoy} and the second term of \eqref{eq:secondboundarytwoy} couple the second site to the giant graviton.
The total contribution from these four terms is
\begin{equation}
N^2\lp \frac{\lambda}{\sqrt{N}}[a_1,a_1^\dag](a_2 - a_1) + (a_2^\dag - a_1)[a_1,a_1^\dag]\frac{\lambda^*}{\sqrt{N}}\rp
\end{equation}

Lastly the contribution from the two derivative terms in $D_2$ is just a scalar multiple of the one loop result
\begin{equation}
2N^2\lp a_1^\dag - \frac{\lambda}{\sqrt{N}}\rp\lp a_1 - \frac{\lambda^*}{\sqrt{N}}\rp
\end{equation}

For the closed spin chain we were able to complete various squares by inserting two copies of the identity into the quadratic terms, one for each site involved.
Here we can not do such a manipulation immediately because one of the two objects involved is not a spin site: it is a giant graviton.
We can insert the identity for site one $a^\dag_1 a_1 + [a_1, a^\dag_1]$. This manipulation only completes the square in the commutator term coupling sites one and two, and the giant graviton.
We are still left with an incomplete quartic and some quadratic terms.
\begin{equation}
\label{eq:awkward2}
\lp a_1^\dag - \frac{\lambda}{\sqrt{N}}\rp\lb a^\dag_1 a + \lp -\frac{\lambda}{\sqrt{N}}\rp a_1 + a_1^\dag\lp-\frac{\lambda^*}{\sqrt{N}}\rp\rb\lp a_1 - \frac{\lambda^*}{\sqrt{N}}\rp + \lp a_1^\dag - \frac{\lambda}{\sqrt{N}}\rp\lp a_1 - \frac{\lambda^*}{\sqrt{N}}\rp
\end{equation}

The reason we do not have the nice form of the closed spin chain at this point is that we forgot about the contributions of certain non-planar terms.
Indeed, since the $Z$ derivatives do not necessarily have to act on a site, we can loosen the requirement that derivatives must sit next to each other.
There are four such terms in $D_2$ that have non-vanishing contribution:
\begin{equation}
\label{eq:nonplanar}
\mathop{:}\Tr\lp -\partial_{Z}Y_{12}\tilde{Z}\partial_{Y_{12}}\partial_{Z}Z + \partial_{Z}Y_{12}\tilde{Z}\partial_{\tilde{Z}}\partial_{Y_{12}}Z + \partial_{Z}ZY_{12}\partial_{Y_{12}}\partial_{Z}Z - \partial_{Z}ZY_{12}\partial_{\tilde{Z}}\partial_{Y_{12}}Z\rp\mathop{:}
\end{equation}
Consider the action of the first term on the spin chain \eqref{eq:openspinchainschematic}.
The relevant terms are
\begin{align}
\label{eq:twoderiv1}
& - \det\lp Z - \lambda \rp\Tr\lp \frac{1}{Z - \lambda}\frac{1}{Z - \lambda}\frac{1}{Z - \lambda}ZY_{12}\tilde{Z}^{n_1+1}WX_{21} \rp \\
\label{eq:twoderiv2}
& + \det\lp Z - \lambda \rp\Tr\lp \frac{Z}{Z - \lambda} \rp\Tr\lp \frac{1}{Z - \lambda}\frac{1}{Z - \lambda}Y_{12}\tilde{Z}^{n_1+1}WX_{21} \rp \\
\label{eq:twoderiv3}
& + \det\lp Z - \lambda \rp\Tr\lp \frac{1}{Z - \lambda} \rp\Tr\lp \frac{1}{Z - \lambda}\frac{1}{Z - \lambda}ZY_{12}\tilde{Z}^{n_1+1}WX_{21} \rp \\
\label{eq:twoderiv4}
& + \det\lp Z - \lambda \rp\Tr\lp \frac{1}{Z - \lambda}\frac{1}{Z - \lambda} \rp\Tr\lp \frac{1}{Z - \lambda}ZY_{12}\tilde{Z}^{n_1+1}WX_{21} \rp \\
\label{eq:twoderiv5}
& - \det\lp Z - \lambda \rp\Tr\lp \frac{1}{Z - \lambda}\frac{1}{Z - \lambda}\frac{1}{Z - \lambda}ZY_{12}\tilde{Z}^{n_1+1}WX_{21} \rp \\
\label{eq:twoderiv6}
& - \det\lp Z - \lambda \rp\Tr\lp \frac{Z}{Z - \lambda} \rp\Tr\lp \frac{1}{Z - \lambda}Y_{12}\tilde{Z}^{n_1+1}WX_{21} \rp
\end{align}
where we have not made use of the expansion $Z = Z - \lambda + \lambda$.
Making use of the expansion, we see that equations \eqref{eq:twoderiv1}, \eqref{eq:twoderiv3}, \eqref{eq:twoderiv4}, and \eqref{eq:twoderiv5} pick up a factor of $\lambda$.
The factor of $1$ from expanding $Z / (Z - \lambda)$ usually ends up reducing at which order of $N$ the corresponding term contributes.
In \eqref{eq:twoderiv2} and \eqref{eq:twoderiv6}, however, we have a trace over the $1$ and thus pick up a factor of $N$.
These terms come in at order $N^2$ but are only quadratic in $\lambda$ and Cuntz operators.
The remaining six terms contain triple pole terms which all combine to give a single second derivative with respect to $\lambda$.
The contribution to the Hamiltonian from \eqref{eq:nonplanar} is
\begin{equation}
N^2\lp a_1^\dag - \frac{\lambda}{\sqrt{N}}\rp \lp -\frac{\lambda}{\sqrt{N}}\rp\lp -\frac{\lambda^*}{\sqrt{N}}\rp \lp a_1 - \frac{\lambda^*}{\sqrt{N}}\rp - N^2 \lp a_1^\dag - \frac{\lambda}{\sqrt{N}}\rp\lp a_1 - \frac{\lambda^*}{\sqrt{N}}\rp
\end{equation}
Note that the sign on the quadratic term is exactly opposite that of \eqref{eq:awkward2} providing a nice cancellation.

The final result after summing everything up and putting back the numerical factors is
\begin{align}
H_2 &=  -\frac{1}{8}\lp\frac{g_{YM}^2N}{4\pi^2}\rp^2 \left[ \lp \frac{\lambda}{\sqrt{N}} - a_1^\dag\rp^2\lp \frac{\lambda^*}{\sqrt{N}} - a_1\rp^2 + \sum_{i=1}^{k-1}(a_{i+1}^\dag - a_i^\dag)^2(a_{i+1} - a_i)^2 \right. \nonumber \\
&\quad + \lp a_k^\dag - \frac{\tilde{\lambda}}{\sqrt{N}}\rp^2\lp a_k - \frac{\tilde{\lambda}^*}{\sqrt{N}}\rp^2 + \lp \frac{\lambda}{\sqrt{N}} - 2a_1^\dag + a_2^\dag\rp[a_1,a_1^\dag]\lp \frac{\lambda^*}{\sqrt{N}} - 2a_1 + a_2\rp\nonumber \\
&\quad + \sum_{i=2}^{k - 1} (a^\dag_{i+1} - 2a_i^\dag + a_{i-1}^\dag)[a_{i},a^\dag_{i}](a_{i+1} - 2a_i + a_{i-1}) \nonumber \\
&\quad + \left. \lp a_{k-1}^\dag - 2a_k^\dag + \frac{\tilde{\lambda}}{\sqrt{N}}\rp[a_k, a_k^\dag]\lp a_{k-1} - 2a_k + \frac{\tilde{\lambda}^*}{\sqrt{N}}\rp \right] \label{eq:twoloopcuntz}
\end{align}
In the coherent state formalism, the commutator yields a factor $(1 - |z_i|^2)$ and so the Hamiltonian becomes
\begin{align}
\label{eq:twoloopcoherent}
H_2 &\rightarrow -\frac{1}{8}\lp\frac{g_{YM}^2N}{4\pi^2}\rp^2\lb \sum_{i=0}^{k}|z_{i+1} - z_i|^4 + \sum_{i=1}^{k} |z_{i+1} - 2z_i + z_{i-1}|^2(1 - |z_i|^2)\rb
\end{align}
where we have taken $z_0 = \lambda^* / \sqrt{N}$ and $z_{k+1} = \tilde{\lambda}^* / \sqrt{N}$.

We notice that the one loop Hamiltonian is a sum of  squares of discrete first derivative operators, whereas the two loop Hamiltonian is the sum of discrete first derivative terms raised to the fourth power plus a  square of discrete second derivative  terms, except with the commutator $[a_i, a_i^\dag]$ inserted in the center.
This commutator projects onto spin chains with sites with zero occupation number.
Thus these terms should not be important at high occupation number.
In the coherent state formalism, large occupation number corresponds to $|z|$ being close to $1$.

Next we look at the ground state.
The first term in \eqref{eq:twoloopcoherent} has the same minimum in the $z_i$ as that at one loop order; the requirement is that $z_1 - z_0 = z_2 - z_1 \dots$.
At this minimum the second term of \eqref{eq:twoloopcoherent} vanishes. Since this term is self-adjoint and positive, the ground state is an eigenstate of this part of the two-loop Hamiltonian and in the end it is an eigenstate of the full two-loop Hamiltonian. 
Indeed, we find that the ground state wave function is the same as before.
This suggests that there is a non-renormalization theorem at play.
The ground state energy changes by a factor of 
\begin{equation}
E_0^{(2)} = -\frac{1}{8}\lp\frac{g_{YM}^2N}{4\pi^2}\rp^2\frac{1}{(k+1)^3}|z_{k+1} -z_0 |^4 = -\frac{1}{8}\lp\frac{g_{YM}^2N}{4\pi^2}\rp^2\frac{1}{(k+1)^3}|\tilde \xi - \xi |^4 \label{eq:secorder}
\end{equation}
which is again a function only of $\tilde\xi -\xi$, which we have argued is our candidate for a central charge for open strings stretching between two D-branes. 
Again, the superscript in the energy indicates that the result is to two loop order.
Hence we have found evidence that the energy of the ground state is controlled only by the complex number $\xi -\tilde \xi$, which gives further evidence for this identification.
This is also in accord with the giant magnon picture \cite{Hofman:2006xt}.

\section{Higher Loops and a Relativistic Dispersion Relation}
\label{sec:lorentz}

So far, we have done our calculations without trying to understand the interpretation of the results from the point of view of $AdS_5\times S^5$ in detail. The purpose of this section is to do this and to use the $AdS_5\times S^5$ geometric intuition to conjecture how the higher loop order corrections might look like to all orders in perturbation theory in detail. 
There are two important things we need to consider. 
First, we saw evidence for being able to measure a central charge for open strings in terms of a difference of coordinates of the end-points of the strings. We need to try to understand the significance of this observation on $AdS_5\times S^5$. The existence of a central charge suggests that there is a BPS formula that would characterize the answer for the all loop result. The second thing we need to do is to try to understand how our perturbative results relate to local ten dimensional  physics on $AdS_5\times S^5$. Indeed, this is the obvious starting point for studying D-branes in $AdS_5\times S^5$, but since in the dual field theory the geometric concepts are emergent, we need to ask ourselves, exactly, what is emerging from our answers?

The first thing we will do therefore is to make an educated guess for the answer of the energy of the ground state of the string stretching between two such giants. Our first observation is that when we take the D-brane to the edge of the BPS quantum droplet, we seem to recover the dispersion relation for giant magnons \cite{Beisert:2005tm} and their bound states \cite{Dorey:2006dq}. Indeed, such an expression would be of the form
\begin{equation}
\Delta - J_1 = \sqrt{(k+1)^2+ \frac{\kappa}{\pi^2} \sin^2 \left(\frac p 2 \right) }
\end{equation} 
where we have a bound state of $k+1$ constituents, and the BMN momentum is $p$, and the 't Hooft coupling is identified as $\kappa= g_{YM}^2 N$.
The formula for the energy of the giant magnon relation depends on having a centrally extended $SU(2|2)$ symmetry on the worldsheet for which the giant magnon produces a short representation.
Although in principle one could have a more general function of $\kappa$ appearing in the square root, this form matches both the weak coupling expansion, as well as the $AdS_5\times S^5$ sigma model limit \cite{Hofman:2006xt}. This suggests that this is the exact formula for all 
$\kappa$, suggesting a particular non-renormalization theorem. One can make other field theory arguments that suggest that this is the only result that is compatible with S-duality \cite{Berenstein:2009qd}

In the case where we send  $\xi,~\tilde{\xi}$ towards the boundary, we have that both become unitary
$\xi = \exp(i p_1)$, $\tilde{\xi} = \exp(ip_2)$, and then 
\begin{align}
\xi - \tilde{\xi} &= 2i e^{i(p_1 + p_2)/2}\frac{e^{i(p_1 - p_2)/2} - e^{-i(p_1 - p_2)/2}}{2i} \\
&= 2i e^{i(p_1 + p_2) / 2} \sin\left(\frac {\Delta p_{12}}{2} \right)
\end{align}
so that
\begin{align}
|\xi-\tilde{\xi}|^2 = 4 \sin^2\left(\frac {\Delta p_{12}}{2} \right)
\end{align}
If we identify $\Delta p_{12}= p$ and we assume compatibility with the giant magnon dispersion relation we conclude that in this limit
\begin{align}
\label{eq:conjecture}
\Delta - J_1 = \sqrt{(k+1)^2+ \frac{g_{YM}^2 N}{4\pi^2} |\xi - \tilde{\xi}|^2}
\end{align}
would give the correct energy of the string to all orders in perturbation theory. We notice that our results in equations \eqref{eq:ineq} (properly normalized) and \eqref{eq:secorder} plus the tree level result $E^{(0)}_0=k+1$ together give the three first terms of the expansion of this formula in the Taylor  series expansion in $\kappa= g_{YM}^2 N$ around $ \kappa=0$. 
Indeed, if $\xi$ and $\tilde \xi$ compute the central charge associated to each end-point of the string, then we must conclude that this result should be valid for all $\xi$ and not just for those special $\xi$ that are unitary
and live at the edge of the quantum droplet. To test this conjecture, let us look at the special case $\xi=\tilde \xi$. This would indicate that the two giant gravitons are on top of each other. As such, these excitations stretching between the two giants should have the same spectrum as the excitations of a giant to itself. After all, when the D-branes are on top of each other, there is an enhanced (gauge) symmetry of coincident 
D-branes \cite{Dai:1989ua}. Such a dispersion relation would give us $ \Delta-J_1= k+1$ which indeed saturates a BPS inequality. Such a state would belong to the chiral ring of the $\mathcal{N}=4 $ SYM theory and can be compared to the DBI fluctuations of giant gravitons finding an exact match \cite{Das:2000st}. Indeed, the spectrum of such fluctuations is independent of the size of the giant graviton and here we find a match to those results. 

One can do further progress with this. The giant graviton background is made of $Z$ fields, and this background breaks the $SO(6)$ R-symmetry to an $SO(4)$ unbroken subgroup. The $Y$ can be considered the highest weight of a vector representation of this unbroken $SO(4)$. This $SO(6)$ R-symmetry is related to the isometry  rotations of the $S^5$ into itself. The $SO(4)$ is the little group unbroken by the giant graviton, and it performs rotations of the giant graviton into itself (this is an $S^3_{GG} \subset S^5$ rotating in $S^5$ \cite{McGreevy:2000cw}). We identify the $S_{GG}^3$ of the giant graviton as a different $S^3$ than the one that determines the radial quantization of the $\mathcal{N}=4$ SYM theory where the computations are made.
 A state with the quantum numbers of $Y^{(k+1)}$ has angular momentum $k+1$ along this $SO(4)$, and thus we should think of our string ground state as a state with angular momentum $k+1$ (this is in the same way that momentum is carried for closed strings \cite{BMN}). 

Now, let us assume that the conjecture \eqref{eq:conjecture} does indeed capture the full result to all orders for the lightest string with angular momentum $k$ stretching between two giant gravitons. The result is a square root and this suggests that we interpret it as a relativistic dispersion relation for a massive particle in curved space. 

Let us consider first the case of branes in flat space. When two D3-branes come close to each other, the low energy effective field theory on their worldvolume is ${\cal N}=4 $ SYM on the Coulomb branch \cite{Polchinski:1995mt} \footnote{This is because the ground state D-brane configuration breaks half of the supersymmetries of flat space, and the only low energy field theory with $16$ supersymmetries and spin content with spin less than or equal to $1$ is the $\mathcal{N}=4 $ SYM itself.}. In the Coulomb branch, the vacuum expectation values of configurations that describe vacua consist of commuting matrices, which can be diagonalized. The positions of the D-branes are the eigenvalues themselves. Given the positions of two D-branes $\vec X_1$ and $\vec X_2$ in the transverse direction to the branes, the $W$ bosons have masses proportional to $|\vec X_1-\vec X_2|$ and these are not renormalized: the massive vector multiplets of $\mathcal{N}=4$ SYM are short representations with a central charge proportional to $\vec X_1-\vec X_2$ itself.  This intuition should apply also if we embed the D-branes in a curved manifold and we make the D-branes parallel to each other (in a curved manifold, where the two D-branes are BPS states we take this to mean that the shortest distance between the branes can be computed anywhere and the results don't depend on where we do this due to a group symmetry).

In the short distance limit between the branes, the same intuition should hold, because we should be able to take a low energy field theory limit where the masses and the geometry are fixed, but the string scale is taken to the $\alpha'\to 0$ limit, just as in the seminal AdS/CFT paper\cite{Maldacena:1997re}. If there is a notion of a position transverse to the brane, so that $\vec X_1-\vec X_2$ makes sense, the mass of a $W$ boson should be proportional to the distance between the branes. The effective distance can be computed using the central charge. If we have a constant mass for a $W$ boson independent of the position along the brane, then the problem of computing the spectrum of $W$ bosons should reduce at low energies to the problem of computing a relativistic dispersion relation for $\mathcal{N}=4$ SYM on the Coulomb branch in curved space. In this calculation the curved space is along the worldvolume of the nearby branes themselves. These $W$ bosons can be of spin one or zero depending on the details of the state, but they will all belong to the same representation of supersymmetry in flat 10 dimensional space.

In our example, the central charge is $|\xi -\tilde \xi|$, and the worldvolume of the D-brane is a curved $S_{GG}^3$, the worldvolume of the giant gravitons themselves. In the limit $\xi-\tilde \xi=0$, the $Y$ fluctuations can be thought of as changing the orientation of the brane embedding into the $S^5$, so it is natural to think of them as affecting the goldstone modes that result from breaking $SO(6)$ down to $SO(4)$ on the D-brane worldvolume. As such, when we turn on the separation between the branes it makes sense to identify the $W$ boson states we get as those that arise from Goldstone bosons in the presence of spontaneous gauge symmetry breaking. Since these are eaten up by the longitudinal component of the massive $W$ bosons, this suggests that the states for which we have computed the mass are part of the massive vector particle, rather than a scalar particle in the $W$ multiplet. Indeed, if we identify the number of $Y$ as a momentum, we see that there is no obvious string ground state at zero momentum: in that case there is no $Y$ connecting the two giants. This is expected, as fluctuations of the Goldstone boson at zero momentum can be gauged away. This suggests that out of the $k+1$ $Y$ fields, only $k$ of them should be counted as momentum, and the last one should come from the spin of the $W$ particle.
 
Indeed, consider a free conformal field theory in four dimensions compactified on a sphere of radius $R$ times time. If such a field theory has a free scalar field, the scalar field will couple to the background curvature of the sphere with a non-minimal conformal coupling. This is, one would need to solve for the spectrum of the second order differential operator
\begin{equation}
\partial_t^2 - \nabla_{S^3}^2 - a R^{-2}
\end{equation}
where $a R^{-2}$ is the term with the Ricci scalar of the background metric.

The energy levels of such a conformally coupled scalar on the sphere will be
\begin{equation}
E_\ell = \frac{\ell +1}R
\end{equation}
starting at $\ell =0, \dots$, where $\ell$ is the principal quantum number for spherical harmonics on the sphere. For a massive scalar field of mass $m$, with a conformal coupling, we would instead derive that 
\begin{equation}
E_\ell=\sqrt{\left(\frac{\ell +1}R\right)^2 + m^2 }
\end{equation}
If we set $R=1$, we get that 
\begin{equation}
E_\ell=\sqrt{\left({\ell +1}\right)^2 + m^2 }
\end{equation}
Which shows that the effective laplacian gets a shift that makes it into a square. We have found a similar equation, but we would need to identify $k+1\to \ell$, whereas we seem to be getting instead $k+1\to \ell+1$. Such a difference can be accounted by a unit of spin.

Notice that we are in better shape than this argument suggests. Indeed, the $AdS_5\times S^5$ and $\mathcal{N}=4$ SYM on $S^3$ themselves have $32$ supersymmetries. The superconformal group $SU(2,2|4)$ admits no central extension. The only way that we get a central extension to have non-zero values in $\mathcal{N}=4$ SYM is by spontaneously breaking the scaling symmetry, but keeping the flat space supersymmetry. Indeed, the central charge extension is necessary to keep the spin to be smaller than or equal to one.
Only half the supersymmetry of the original system survives when we do this, and this is done by going to the Coulomb branch. 

By the same token, the presence of the giant gravitons breaks the conformal symmetry of $SU(2,2|4)$ so that only half of the supersymmetries are unbroken. Given that supersymmetry was broken to half, one can now
argue that a central charge extension can appear for the open strings stretching between giants in a similar way to what happens in flat space. Moreover, there are $16$ supersymmetries acting on the system that do not
act on the D-brane system, which is considered a ground state. To only have particles of spin less than or equal to one survive as $W$ bosons in the presence of so many supersymmetries, it must be the case that there is a central charge extension. Otherwise one would have long representations of supersymmetry (they would have $2^8$ states, rather than $2^4$) which include particles of spin higher than one and this would be inconsistent with the expectations of low energy physics of D-branes. All of the other states in the multiplet should therefore be accessible by acting with the unbroken supersymmetries. Some of these commute with the twisted Hamiltonian $H_{BMN}=\Delta -J_1$ and should produce degeneracies with other states that have different spin. Thus, in a formula like \eqref{eq:conjecture}, the $SO(4)$ charge could change by one unit, but the spin could also change in such a way that a (massive) particle of spin zero on the $S^3$ has the same energy as a vector particle with the right polarization, but where the splitting of quantum numbers into momentum versus spin is different.  

Further evidence for this identification of the coordinates of the quantum hall droplet as giving rise to a central charge comes from considering the so called dual giant gravitons (those that grow into $AdS$ rather than on the $S^5$ sphere \cite{Grisaru:2000zn,Hashimoto:2000zp}). In the quantum hall droplet picture, such states are interpreted as an eigenvalue of the $Z$ matrix acquiring a large expectation value and spontaneously breaking the original $U(N)$ gauge symmetry of the $\mathcal{N}=4$ SYM to $U(N-1)\times U(1)$ \cite{Hashimoto:2000zp,Berenstein:2004kk}, where this is a constant field configuration on the $S^3$ of the original theory, and the dual giant is also of the shape of an $S^3$, which now is `parallel' to the boundary $S^3$.
The classical configurations of the $Z$ that satisfy the corresponding BPS conditions are exactly points on the Coulomb branch of $\mathcal{N}=4$ SYM, and one can extend this idea to $1/8$ BPS states \cite{Berenstein:2005aa}. Indeed, this idea that the configurations on moduli space can be turned to dual giants is applicable for fairly general $AdS\times X$ geometries and one can also argue that this is enough to reproduce plane wave limit spectra and supergravity spectra from field theory \cite{Berenstein:2007wi,Berenstein:2007kq}. For this setup, the eigenvalues of $Z$ themselves determine the position of the branes on the Coulomb branch and serve as the $X$ coordinates, as well as the central charge. If we take a scaling limit of large eigenvalues, the effective mass of all the $W$ bosons can be made arbitrarily large, much larger than the scale of compactification of the original field theory on the $S^3$, so that a flat space limit can be taken by going to intermediate energies (momentum of order the mass, which correspond to wavelengths much shorter than the $S^3$ inverse radius) and we can really think of the system as $\mathcal{N}=4$ SYM on the Coulomb branch in flat space. This is seen also by taking appropriate limits in  supergravity solutions \cite{LLM}.

Perhaps a more convincing argument is to follow the work of \cite{Beisert:2005tm,Beisert:2006qh} more closely. In that work Beisert argues that the central charge on the spin chain with a ferromagnetic $Z$ background corresponds to adding (or subtracting) a $Z$ at the left and right of asymptotic excitations (this only works on the infinite chain limit). In our case, we would want to produce such a central charge extension for a finite open string in such a way that it is compatible with the asymptotic prescription.  On each excitation we want to replace $Y\to [Y,\partial_Z]$, or $Y\to [Z,Y]$, just like in the infinite chain so that we can add or  subtract a $Z$ from the chain around each $Y$. This is realized by the Cuntz chain operators $\sqrt{N} (a_i^\dagger - a_{i+1}^\dagger)$, or $(a_i-a_{i+1})/\sqrt{N}$ (the additional normalization factor of $\sqrt{N}$ is due to the change in the norm of the state with a different length of the spin chain). To add a $Z$ to the left of the chain, in our notation for operators, we would use 
the identity $( Z-\lambda)^{-1}Z= 1+ \lambda(Z-\lambda)^{-1}$ to show that we get a factor of $\lambda$ (the term with the one would be non-planar and would remove the boundary of the string on the left, joining it with another string). Thus we would get that the asymptotic central charge for each $Y$ is either $\sqrt{N} (a_i^\dagger - a_{i+1}^\dagger)$ or $\lambda- \sqrt N a_1^\dagger$ and a similar term for the right boundary. For the case where we subtract a $Z$, we would identify $\xi-a_1, a_1-a_2, \dots a_{k}-\tilde \xi$ as the values of the central charge for each $Y$ defect. The total central charge is then 
\begin{equation}
\lambda- \sqrt N a_1^\dagger+\sum_i (\sqrt{N} (a_i^\dagger - a_{i+1}^\dagger)) +\sqrt N a_k-\tilde \lambda = \lambda-\tilde \lambda
\end{equation}
 Obviously, we see then that the condition to get a short multiplet of the centrally extended $SU(2|2)$ would correspond to the equation \eqref{eq:conjecture} exactly, and the identification of $\xi-\tilde\xi$ as the central charge of the full string is inevitable.

All our arguments so far are in order to make the claim that equation \eqref{eq:conjecture} is correct to all orders in perturbation theory. Notice that we are claiming that this equation should be interpreted as a relativistic dispersion relation for a (local) field theory on $S^3_{GG}$, the worldvolume of the giant gravitons themselves. The fact that we can reproduce this to second order in $\lambda$ suggests that we are actually probing properties associated to local Lorentz invariance in higher dimensions. Notice that from the boundary field theory it is natural to assume that time and the original $S^3$ are related by locality and causality, but that does not make it automatic for $S^3_{GG}$, since this $S^3_{GG}\subset S^5$ is emergent itself. Indeed, the new local lorentzian structure would mix an $SO(4)_R$ symmetry with time, rather than the $SO(4)\subset SU(2,2)\simeq SO(4,2)$ which is also unbroken by the giant graviton. Notice also that this relativistic dispersion relation is compatible with the usual way of thinking about the Higgs mechanism arising from D-branes in string theory.

A more pressing question is if we can also make a conjecture to extend our effective Hamiltonian given by \eqref{eq:twoloopcuntz} to all orders, rather than just the energy of the ground state. It is interesting to phrase this in terms of the effective Hamiltonian for a coherent state as in equation \eqref{eq:twoloopcoherent}. As seen previously, the ground state of the two loop Hamiltonian and the one loop Hamiltonian agree, so that $z_i-z_{i+1}= z_{i-1}- z_i$. The structure and coefficients of the terms that only involve nearest neighbors suggest that a part of the Hamiltonian is given by
\begin{equation}
H_{\text{nn}} = \sum_{i=0}^{k} \sqrt{1+ \frac{\kappa}{4 \pi^2} |z_{i+1}-z_i|^2} \label{eq:BPSsum}
\end{equation}
again with $z_0=\xi$ and $z_{k+1} =\tilde \xi$. On expanding this expression terms of Cuntz oscillators, we should replace $z_i \to a_i$, $\bar z_i \to a_i^\dagger$ for $i=1, \dots, k$ and require normal ordering, with all the daggered objects appearing to the left of the undaggered objects.

If we assume that the ground state is not renormalized, so that $(k+1)(z_i -z_{i+1}) = (\xi-\tilde \xi)$, we find that the energy of the string would completely be accounted for by these terms as follows
\begin{equation}
H_{\text{nn}}  |\psi_0\rangle =  (k+1) \sqrt{1+ \frac{\kappa}{4 \pi^2(k+1)^2} |\xi-\tilde \xi|^2} = \sqrt{(k+1)^2+ \frac{\kappa}{4\pi^2} |\xi-\tilde \xi|^2 }
\end{equation}
The non-nearest neighbor interactions should contain terms like the ones in \eqref{eq:twoloopcoherent} that seem to be related to discretized versions of higher derivatives, multiplied by factors of $(1-|z_i|^2)$. These factors are easy to explain in perturbation theory: they are the probabilities that two or more $Y$ do not have any $Z$ between them (after all, they appear from projectors onto the ground state $P_0$), so that they correspond to jumping over more than one $Y$. Such terms are required by the $SU(2)$ symmetry of the spin chain. If these terms always show up as discretized versions of higher derivatives that cancel for the ground state, this suggests that the $\ket \psi_0$ is a common eigenvalue for a large number of operators appearing in the Hamiltonian. We think this should be related to being protected by supersymmetry. Notice also that these terms vanish exactly when we send $|z_i|\to 1$. This is the limit of large occupation number, where the $Y$ defects would never be able to cross each other, and where one would match with the asymptotic Bethe ansatz of the infinite closed chain in a natural way. In such a setup, we must also get that the Hamiltonian asymptotes to \eqref{eq:BPSsum}, as this is what one expects from bound states of magnons because there the BMN momentum can be complexified \cite{Dorey:2006dq}, and we already have our identification of momentum and the central charge. The fact that all these extra higher derivative operators are zero at first order suggests that we can start with the Hamiltonian which only has nearest neighbor interactions and use these extra terms as perturbations. In such a Hamiltonian the effect between nearest neighbors is identical to what we would have if we were stretching strings between D-branes at the $z_i$. The extra terms that correspond to higher derivatives would lower the energy if the string is not made of straight edges (at least the first one such term does), so they should account for binding energies between these string bits. Such a geometric picture is compatible with the ideas of \cite{Berenstein:2005aa,Berenstein:2005jq} and makes them more precise.

Notice that surprisingly, this also matches the result of \cite{Hatsuda:2006ty}, not just at the level of the formula, but the geometry of the $z$ themselves.  In \cite{Hatsuda:2006ty} the identification of the $z_i$ with the droplet coordinates of ``D-branes'' as eigenvalues is performed assuming the general strategy of gases of eigenvalues is correct. Notice that the number of such eigenvalues needed is equal to the number of $Y$ plus one. In our case, the two external ones should be thought of as true D-branes, whereas the ones associated to the other $z_i$ should be more virtual in character, as they have not been removed from the droplet before we put the string there. 

Notice also that having some of the momentum of the string being proportional to the number of D-branes is common in matrix theory \cite{BFSS}. This might also suggest alternative approaches to understand if there is a plane wave limit matrix model for type IIB strings \cite{Verlinde:2002ig} and perhaps might realize some of these ideas in more detail. Similarly, this might lead to a possible derivation of the tiny graviton matrix model
\cite{SheikhJabbari:2004ik}, as now the spectrum of strings between giant gravitons in the plane wave limit can be within reach of calculations in the ${\cal N}=4$ SYM itself. 

Having this relativistic dispersion relation has consequences for our understanding of locality on the sphere in the directions transverse to the D-brane, not only along its worldvolume.
 So long as the gap in anomalous dimensions is typically large, taking $k$ large but finite and varying $\lambda -\tilde\lambda$ to be small enough so that $g_{YM}^2N |\lambda -\tilde\lambda| \simeq k^2$, one can see that 
 at large t'Hooft coupling one can corner oneself so that $|\xi -\tilde \xi|<<1$ which means we have probed the geometry of the sphere on distances much smaller than the AdS radius $R$ transversely to the D-branes, and on distances of order $R/k$ along the giant graviton worldvolume. The limit where both terms are of the same size is of the order of the Compton wavelength of the string as a W-boson in the directions along the D-brane. This is much smaller than the AdS radius, so it provides evidence for locality on much smaller scales than the AdS radius in all directions along the sphere. One can also argue that this extends to the AdS geometry itself: after all, the giant gravitons are localized at the origin of AdS in global coordinates.
In principle we can move them so that they can be at different positions in the AdS radial direction by giving the giant gravitons some angular momentum (these are the result of superconformal transformation on the different giant graviton states). Presumably one can find evidence in such a case that the corresponding W bosons would also get a mass from the AdS displacement which would also indicate local physics along all of the AdS directions together with the sphere.
 
One can also consider the so called dual giant gravitons, which grow into AdS but are point like on the sphere.  The dual states to those D-branes growing into the AdS directions are known but not their precise collective coordinate description in the full quantum theory.  To zeroth order one thinks of them as spontaneous symmetry breaking of the original $U(N)$ gauge field theory to $U(N-1)\times U(1)$ by a particular time dependent vev. 
For that case, the mass of W bosons is determined from the classical physics of this symmetry breaking. One should then realize these states in the geometry, as branes in the free fermion droplet \cite{Berenstein:2004kk,LLM}. Since these states break the same symmetries of the conformal group as the ordinary giant gravitons, one could expect that the string attached to them would also measure a central charge which is measured by the coordinate of the giant graviton in the quantum hall droplet plane geometry. The details of such a calculation are beyond the scope of the present paper.

\section{Conclusion}

In this paper we began by considering a class of spin chain operators in the $SU(2)$ subsector of $\mathcal{N} = 4$ SYM with a fixed number of $Y$'s and an arbitrary number of $Z$'s between them,
with D-branes inserted at the ends so as to form open string states in the CFT. These D-branes are parametrized by a collective coordinate.
The spin   chain  states were described by a Cuntz oscillator basis and to solve the system we also introduced a collective coordinate approach using coherent spin chain states for the Cuntz oscillator themselves.
We found that the one loop Hamiltonian contains self energy and hopping terms which can be written as a sum of squares. It is this property of having a spin chain Hamiltonian as being  a sum of squares that suggests that properly thinking about the ground state can be interpreted as some sort of BPS constraint.
The open spin chain modifies the one loop Hamiltonian through boundary terms which amount to adding $c$-number sites to each end of the spin chain whose values are the normalized collective coordinates of the giant gravitons. The ground state for the open spin chain is characterized by linearly interpolating the spin chain collective coordinates in the sites between the giant graviton collective coordinates that act as end points.

At the next  loop order we found contributions quartic in discrete derivatives between nearest neighbor sites with additional terms coupling neighbors to next to nearest neighbors that mimicked a discrete version of a second derivative.
However, these terms also contain projections onto the zero occupation states causing them to contribute to the Hamiltonian only when acting on sites with low occupation number. 
The open spin chain modifies the two loop Hamiltonian on the boundary exactly how it modified the hamiltonian in the bulk. Furthermore we found that the ground state of the two loop spin chain was identical to the ground state of the one loop spin chain, and that the two loop energy gave further evidence for the central charge extension property of the open spin chain. Surprisingly the next to nearest neighbor contribution to the energy vanished when acting on the ground state of the spin chain, not just as a vev, but as an operator. The two loop contribution to the total energy happens in a way that is compatible with a relativistic dispersion relation for the open string to second order in an expansion of a dispersion relation of the form $E=\sqrt{p^2 + m^2}$ in a Taylor series about $m^2=0$, with the precise coefficients in place. We conjectured that the higher loop contributions complete the square root series and that the ground state wave function is not renormalized. Our evidence was very strong, as we showed that it follows from an argument involving a central charge extension of the unbroken supersymmetry algebra that leaves the giant gravitons invariant.

Furthermore, we conjectured that only the nearest neighbor interactions in the higher loop spin chain could contribute to the ground state energy and that all the higher neighbor interactions vanish as operator equations on the ground state of the spin chain. We argue that this suggests an expansion where these higher order discrete derivative terms are used as a perturbation, where the zeroth order contribution includes only the nearest neighbor interactions. Such a picture is very similar to the picture of geometries being built by string bits stretching between D-branes that form a non-trivial quantum gas. The nearest neighbor interactions would be the energies of these string bits, and the higher order terms that are missed would deal with the interactions between these string bits.  Because the expansion seems to be written in terms of multiple discrete derivative operators, one can imagine that in the continuum limit these might fully restore the string sigma model in a particular gauge, where
one can also recover higher derivative $\alpha'$ corrections of the string propagation on $AdS_5\times S^5$. To further understand this issue one necessarily has to go to higher loop orders.

As we have argued in this paper, string states stretched between D-branes give us some insight into how geometry is realized. Although this is not a new point of view, understanding its realization in the AdS/CFT correspondence setup is important because it can lead to a much better understanding as to how classical geometry is replaced by a quantum realization of geometry, or when geometry stops making sense. This is in keeping with the idea that the CFT is actually a definition of quantum gravity.
Following this train of thought we have been able to get a glimpse of how locality in higher dimensions can emerge in the AdS/CFT context from a field theory computation. Particularly important for this question is how the gap in anomalous dimensions is generated between different states: those that remain massless in the supergravity limit, and those that become stringy. In our example, the distance between the branes emerges as a measurement of this gap. Indeed, we have found evidence that in the full $AdS_5\times S^5$ geometry giant graviton D-brane defects break the conformal field theory in such a way that they give rise to a central charge extension of the unbroken SUSY algebra, extending previous ideas of this central charge extension that are related to the integrable structure of the ${\cal N}=4$ spin chain model. The simple interpretation of this central charge is that it is expected from trying to understand the emergent  ${\cal N}=4$ SYM for a stack of D-branes in the Coulomb branch (the D-branes in question are the giant gravitons, and the effective ${\cal N}= 4 $ SYM is the dynamics on their worldvolume). 
A non-zero value for such a central charge  gives in principle  a lower bound for operator dimensions and helps explain further this gap in anomalous dimensions. Furthermore, such a central charge can be thought in some instances to provide a natural notion of position, like in the case of flat D-branes in flat space. 

Furthermore, the fact that D-branes carry non-trivial gauge fields on their worldvolume implies that open string joining and splitting necessarily takes place along the world-volume of the D-brane, which means interactions are local in the transverse directions to the D-brane. This only makes sense so long as we can argue that the D-branes are actually local in some geometry in the first place. Presumably, consistency of locality between different such probes requires that physics is local in the geometry. Since the D-branes can be moved closer or father apart from each other, one can argue that so long as the gap is sufficiently large, one has probed locality on distances longer than a Compton wavelength for a W-boson. These distances are much smaller than an AdS radius and thus one is finding locality on sub-AdS lengths. 

Although we have not studied this process yet, the fact that there is an effective gauge theory on giant graviton states is understood because one can see that there is a Gauss' law constraint for counting string states between giants \cite{Balasubramanian:2004nb} (see also \cite{deMelloKoch:2012ck,Berenstein:2013md}). This is intimately tied to the original gauge invariance of the original ${\cal N}=4$ SYM and generalizes to other field theories. To nail the case of locality one would need to show that interactions are polynomial in momenta. There are setups where constituents can have relativistic dispersion relations, but the interactions do not respect Lorentz invariance, such as in noncommutative field theory, which can appear as limits of string theory (see \cite{Douglas:2001ba} for a review). We believe that understanding the precise role that the central charge plays is crucial to understanding locality in ten dimensions.
Understanding exactly how this central charge extension might control the effective field theory generated on the D-brane would be very interesting.

\section{Acknowledgements}

D. B. would like to thank Y. Asano,  S. Hartnoll, E. Silverstein, M. Staudacher and J. Polchinski for various discussions and comments.

Work  supported in part by DOE under grant DE-FG02-91ER40618.  E. D. is supported by the Department of Energy Office of Science Graduate Fellowship Program (DOE SCGF), made possible in part by the American Recovery and Reinvestment Act of 2009, administered by ORISE-ORAU under contract no. DE-AC05-06OR23100.

\appendix

\section{Conventions and Relations for the Lie Algebra of $U(N)$}
\label{app:conventions}

We use the following conventions for the generators of $U(N)$.
These are the same as those used in \cite{Beisert:2003tq}.
\begin{equation}
\Tr(T^AT^B) = \delta^{AB}, \quad (T^A)^\alpha_{\phantom{\alpha}\beta}(T^A)^\gamma_{\phantom{\gamma}\delta} = \delta^{\alpha}_{\phantom{\alpha}\delta}\delta^{\gamma}_{\phantom{\gamma}\beta}
\end{equation}
where $A,~B$ run from $1$ to $N^2$ and repeated indices indicate summation.
These relations can be used to prove the fusion / fission rules
\begin{equation}
\Tr(AT^A)\Tr(BT^A) = \Tr(AB), \quad \Tr(AT^ABT^A) = \Tr(A)\Tr(B)
\end{equation}
The following relations are relevant to computing the order $N^2$ boundary contribution to the two loop Hamiltonian.
If $W$ is a field in the SYM then we may expand it as $X = X^AT^A$ and its derivative as $\partial_X = T^A\partial_{X^A}$.
Using this fact and the relations
\begin{align}
(\partial_{Z})^a_{\phantom{a}b}\det(Z - \lambda) &= \det(Z - \lambda)\lp \frac{1}{Z - \lambda}\rp^a_{\phantom{a}b} \\
(\partial_{Z})^a_{\phantom{a}b}\lp \frac{1}{Z - \lambda}\rp^c_{\phantom{c}d} &= - \lp \frac{1}{Z - \lambda}\rp^a_{\phantom{a}d}\lp \frac{1}{Z - \lambda}\rp^c_{\phantom{c}b}
\end{align}
we have the following fusion / fission rules
\begin{align}
\Tr(A\partial_Z)\det(Z - \lambda) &= \det(Z - \lambda)\Tr\lp A \frac{1}{Z - \lambda}\rp \\ 
\Tr(A\partial_Z)\Tr\lp B \frac{1}{Z - \lambda}\rp &= -\Tr\lp A \frac{1}{Z - \lambda}B\frac{1}{Z - \lambda}\rp \\ 
\Tr\lp A\partial_Z B \frac{1}{Z - \lambda}\rp &= -\Tr\lp A \frac{1}{Z - \lambda}\rp\Tr\lp B \frac{1}{Z - \lambda}\rp
\end{align}
where it is assumed that $A$ and $B$ are independent of $Z$.

When computing the boundary terms at two loop order, we need to collect terms into a derivative with respect to $\lambda$ so that we then pull down its conjugate.
In order to do this we need the relations
\begin{align}
&\hspace{-0.5cm}\partial_\lambda\lb\det(Z - \lambda)\Tr\lp \frac{1}{Z - \lambda}A\rp\rb = \nonumber \\
&\quad \det(Z - \lambda)\lb \Tr\lp\frac{1}{(Z - \lambda)^2}A\rp - \Tr\lp \frac{1}{Z - \lambda}\rp\Tr\lp \frac{1}{Z - \lambda}A\rp\rb \\
&\hspace{-0.5cm}\partial_\lambda^2\lb \det(Z - \lambda)\Tr\lp \frac{1}{Z - \lambda}A\rp\rb = \nonumber \\
&\quad \det(Z - \lambda)\lb 2\Tr\lp\frac{1}{(Z - \lambda)^3}A\rp - 2\Tr\lp\frac{1}{Z - \lambda}\rp\Tr\lp\frac{1}{(Z - \lambda)^2}A\rp \rr \nonumber \\
&\quad \phantom{\det(Z - \lambda)}\quad \lr - \lb\Tr\lp\frac{1}{(Z - \lambda)^2}\rp - \Tr\lp\frac{1}{Z - \lambda}\rp\Tr\lp\frac{1}{Z - \lambda}\rp\rb\Tr\lp\frac{1}{Z - \lambda}A\rp \rb
\end{align}
where $A$ is a matrix that does not depend on $\lambda$.


\begin{thebibliography} {99}


\bibitem{Maldacena:1997re} 
  J.~M.~Maldacena,
  ``The Large N limit of superconformal field theories and supergravity,''
  Adv.\ Theor.\ Math.\ Phys.\  {\bf 2}, 231 (1998)
  [hep-th/9711200].
  
\bibitem{Gary:2009ae} 
  M.~Gary, S.~B.~Giddings and J.~Penedones,
  ``Local bulk S-matrix elements and CFT singularities,''
  Phys.\ Rev.\ D {\bf 80}, 085005 (2009)
  [arXiv:0903.4437 [hep-th]].
  
  
\bibitem{Fitzpatrick:2011ia} 
  A.~L.~Fitzpatrick, J.~Kaplan, J.~Penedones, S.~Raju and B.~C.~van Rees,
  ``A Natural Language for AdS/CFT Correlators,''
  JHEP {\bf 1111}, 095 (2011)
  [arXiv:1107.1499 [hep-th]].
  
\bibitem{Giddings:1999jq} 
  S.~B.~Giddings,
  ``Flat space scattering and bulk locality in the AdS / CFT correspondence,''
  Phys.\ Rev.\ D {\bf 61}, 106008 (2000)
  [hep-th/9907129].
  
\bibitem{Gary:2009mi} 
  M.~Gary and S.~B.~Giddings,
``The Flat space S-matrix from the AdS/CFT correspondence?,''
  Phys.\ Rev.\ D {\bf 80}, 046008 (2009)
  [arXiv:0904.3544 [hep-th]].

\bibitem{Heemskerk:2009pn} 
  I.~Heemskerk, J.~Penedones, J.~Polchinski and J.~Sully,
  ``Holography from Conformal Field Theory,''
  JHEP {\bf 0910}, 079 (2009)
  [arXiv:0907.0151 [hep-th]].


\bibitem{Fitzpatrick:2010zm} 
  A.~L.~Fitzpatrick, E.~Katz, D.~Poland and D.~Simmons-Duffin,
  ``Effective Conformal Theory and the Flat-Space Limit of AdS,''
  JHEP {\bf 1107}, 023 (2011)
  [arXiv:1007.2412 [hep-th]].






\bibitem{Beisert:2010jr} 
  N.~Beisert, C.~Ahn, L.~F.~Alday, Z.~Bajnok, J.~M.~Drummond, L.~Freyhult, N.~Gromov and R.~A.~Janik {\it et al.},
  ``Review of AdS/CFT Integrability: An Overview,''
  Lett.\ Math.\ Phys.\  {\bf 99}, 3 (2012)
  [arXiv:1012.3982 [hep-th]].

\bibitem{Berenstein:2004ys} 
  D.~Berenstein and S.~A.~Cherkis,
  ``Deformations of {\cal N}=4 SYM and integrable spin chain models,''
  Nucl.\ Phys.\ B {\bf 702}, 49 (2004)
  [hep-th/0405215].


\bibitem{Basu:2011di} 
  P.~Basu and L.~A.~Pando Zayas,
  ``Chaos Rules out Integrability of Strings in AdS$_5 \times T^{1,1}$,''
  Phys.\ Lett.\ B {\bf 700}, 243 (2011)
  [arXiv:1103.4107 [hep-th]].

\bibitem{Berenstein:2005aa} 
  D.~Berenstein,
  ``Large N BPS states and emergent quantum gravity,''
  JHEP {\bf 0601}, 125 (2006)
  [hep-th/0507203].


\bibitem{McGreevy:2000cw} 
  J.~McGreevy, L.~Susskind and N.~Toumbas,
  ``Invasion of the giant gravitons from Anti-de Sitter space,''
  JHEP {\bf 0006}, 008 (2000)
  [hep-th/0003075].




\bibitem{Witten:1998xy} 
  E.~Witten,
  ``Baryons and branes in anti-de Sitter space,''
  JHEP {\bf 9807}, 006 (1998)
  [hep-th/9805112].

\bibitem{Gubser:1998fp} 
  S.~S.~Gubser and I.~R.~Klebanov,
  ``Baryons and domain walls in an {\cal N}=1 superconformal gauge theory,''
  Phys.\ Rev.\ D {\bf 58}, 125025 (1998)
  [hep-th/9808075].


\bibitem{Mikhailov:2000ya} 
  A.~Mikhailov,
  ``Giant gravitons from holomorphic surfaces,''
  JHEP {\bf 0011}, 027 (2000)
  [hep-th/0010206].

\bibitem{Berenstein:2002ke} 
  D.~Berenstein, C.~P.~Herzog and I.~R.~Klebanov,
  ``Baryon spectra and AdS /CFT correspondence,''
  JHEP {\bf 0206}, 047 (2002)
  [hep-th/0202150].

\bibitem{Balasubramanian:2001nh} 
  V.~Balasubramanian, M.~Berkooz, A.~Naqvi and M.~J.~Strassler,
  ``Giant gravitons in conformal field theory,''
  JHEP {\bf 0204}, 034 (2002)
  [hep-th/0107119].


\bibitem{deMelloKoch:2011ci} 
  R.~de Mello Koch, G.~Kemp and S.~Smith,
  ``From Large N Nonplanar Anomalous Dimensions to Open Spring Theory,''
  Phys.\ Lett.\ B {\bf 711}, 398 (2012)
  [arXiv:1111.1058 [hep-th]].

\bibitem{Berenstein:1996xk} 
  D.~Berenstein, R.~Corrado, W.~Fischler, S.~Paban and M.~Rozali,
  ``Virtual D-branes,''
  Phys.\ Lett.\ B {\bf 384}, 93 (1996)
  [hep-th/9605168].

\bibitem{Berenstein:2013md} 
  D.~Berenstein,
  ``Giant gravitons: a collective coordinate approach,''
  arXiv:1301.3519 [hep-th].

\bibitem{Berenstein:2004kk} 
  D.~Berenstein,
  ``A Toy model for the AdS / CFT correspondence,''
  JHEP {\bf 0407}, 018 (2004)
  [hep-th/0403110].
  
\bibitem{LLM} 
  H.~Lin, O.~Lunin and J.~M.~Maldacena,
  ``Bubbling AdS space and 1/2 BPS geometries,''
  JHEP {\bf 0410}, 025 (2004)
  [hep-th/0409174].

\bibitem{Berenstein:2005vf} 
  D.~Berenstein and S.~E.~Vazquez,
  ``Integrable open spin chains from giant gravitons,''
  JHEP {\bf 0506}, 059 (2005)
  [hep-th/0501078].
  
  
  
  

\bibitem{Berenstein:2005fa} 
  D.~Berenstein, D.~H.~Correa and S.~E.~Vazquez,
  ``Quantizing open spin chains with variable length: An Example from giant gravitons,''
  Phys.\ Rev.\ Lett.\  {\bf 95}, 191601 (2005)
  [hep-th/0502172].




\bibitem{deMelloetal} 
  R.~de Mello Koch, J.~Smolic and M.~Smolic,
  ``Giant Gravitons - with Strings Attached (II),''
  JHEP {\bf 0709}, 049 (2007)
  [hep-th/0701067].

\bibitem{Berenstein:2007zf} 
  D.~Berenstein and S.~E.~Vazquez,
  ``Giant magnon bound states from strongly coupled {\cal N}=4 SYM,''
  Phys.\ Rev.\ D {\bf 77}, 026005 (2008)
  [arXiv:0707.4669 [hep-th]].

\bibitem{Beisert:2005tm} 
  N.~Beisert,
  ``The SU(2|2) dynamic S-matrix,''
  Adv.\ Theor.\ Math.\ Phys.\  {\bf 12}, 945 (2008)
  [hep-th/0511082].



\bibitem{MZ} 
  J.~A.~Minahan and K.~Zarembo,
  ``The Bethe ansatz for {\cal N}=4 superYang-Mills,''
  JHEP {\bf 0303}, 013 (2003)
  [hep-th/0212208].

\bibitem{D'Hoker:1998tz} 
  E.~D'Hoker, D.~Z.~Freedman and W.~Skiba,
  ``Field theory tests for correlators in the AdS / CFT correspondence,''
  Phys.\ Rev.\ D {\bf 59}, 045008 (1999)
  [hep-th/9807098].

\bibitem{Beisert:2003tq} 
  N.~Beisert, C.~Kristjansen and M.~Staudacher,
  ``The Dilatation Operator of Conformal {\cal N} = 4 Super Yang-Mills Theory,''
  Nucl.\ Phys.\ B {\bf 664}, 131 (2003)
  [hep-th/0303060].

\bibitem{BMN} 
  D.~E.~Berenstein, J.~M.~Maldacena and H.~S.~Nastase,
  ``Strings in flat space and pp waves from {\cal N}=4 superYang-Mills,''
  JHEP {\bf 0204}, 013 (2002)
  [hep-th/0202021].

\bibitem{Berenstein:2003ah} 
  D.~Berenstein,
  ``Shape and holography: Studies of dual operators to giant gravitons,''
  Nucl.\ Phys.\ B {\bf 675}, 179 (2003)
  [hep-th/0306090].



  

  
\bibitem{Berenstein:2005jq} 
  D.~Berenstein, D.~H.~Correa and S.~E.~Vazquez,
  ``All loop BMN state energies from matrices,''
  JHEP {\bf 0602}, 048 (2006)
  [hep-th/0509015].
  
\bibitem{Hatsuda:2006ty} 
  Y.~Hatsuda and K.~Okamura,
  ``Emergent classical strings from matrix model,''
  JHEP {\bf 0703}, 077 (2007)
  [hep-th/0612269].

\bibitem{Dorey:2006dq} 
  N.~Dorey,
  ``Magnon Bound States and the AdS/CFT Correspondence,''
  J.\ Phys.\ A {\bf 39}, 13119 (2006)
  [hep-th/0604175].

\bibitem{Corley:2001zk} 
  S.~Corley, A.~Jevicki and S.~Ramgoolam,
  ``Exact correlators of giant gravitons from dual N=4 SYM theory,''
  Adv.\ Theor.\ Math.\ Phys.\  {\bf 5}, 809 (2002)
  [hep-th/0111222].

\bibitem{Balasubramanian:2004nb} 
  V.~Balasubramanian, D.~Berenstein, B.~Feng and M.~-x.~Huang,
  ``D-branes in Yang-Mills theory and emergent gauge symmetry,''
  JHEP {\bf 0503}, 006 (2005)
  [hep-th/0411205].

\bibitem{Koch:2010gp} 
  R.~d.~M.~Koch, G.~Mashile and N.~Park,
  ``Emergent Threebrane Lattices,''
  Phys.\ Rev.\ D {\bf 81}, 106009 (2010)
  [arXiv:1004.1108 [hep-th]].


\bibitem{Gross:2002su} 
  D.~J.~Gross, A.~Mikhailov and R.~Roiban,
  ``Operators with large R charge in {\cal N}=4 Yang-Mills theory,''
  Annals Phys.\  {\bf 301}, 31 (2002)
  [hep-th/0205066].

\bibitem{Santambrogio:2002sb} 
  A.~Santambrogio and D.~Zanon,
``Exact anomalous dimensions of {\cal N}=4 Yang-Mills operators with large R charge,''
  Phys.\ Lett.\ B {\bf 545}, 425 (2002)
  [hep-th/0206079].





\bibitem{Dai:1989ua} 
  J.~Dai, R.~G.~Leigh and J.~Polchinski,
  ``New Connections Between String Theories,''
  Mod.\ Phys.\ Lett.\ A {\bf 4}, 2073 (1989).
  
\bibitem{Witten:1978mh} 
  E.~Witten and D.~I.~Olive,
  ``Supersymmetry Algebras That Include Topological Charges,''
  Phys.\ Lett.\ B {\bf 78}, 97 (1978).


\bibitem{Seiberg:1994rs} 
  N.~Seiberg and E.~Witten,
 ``Electric - magnetic duality, monopole condensation, and confinement in N=2 supersymmetric Yang-Mills theory,''
  Nucl.\ Phys.\ B {\bf 426}, 19 (1994)
  [Erratum-ibid.\ B {\bf 430}, 485 (1994)]
  [hep-th/9407087].


\bibitem{Hofman:2006xt} 
  D.~M.~Hofman and J.~M.~Maldacena,
  ``Giant Magnons,''
  J.\ Phys.\ A {\bf 39}, 13095 (2006)
  [hep-th/0604135].

\bibitem{Bena:2003wd} 
  I.~Bena, J.~Polchinski and R.~Roiban,
  ``Hidden symmetries of the AdS(5) x S**5 superstring,''
  Phys.\ Rev.\ D {\bf 69}, 046002 (2004)
  [hep-th/0305116].

\bibitem{Dolan:2003uh} 
  L.~Dolan, C.~R.~Nappi and E.~Witten,
  ``A Relation between approaches to integrability in superconformal Yang-Mills theory,''
  JHEP {\bf 0310}, 017 (2003)
  [hep-th/0308089].


  
  
\bibitem{Berenstein:2006qk} 
  D.~Berenstein, D.~H.~Correa and S.~E.~Vazquez,
  ``A Study of open strings ending on giant gravitons, spin chains and integrability,''
  JHEP {\bf 0609}, 065 (2006)
  [hep-th/0604123].
  
\bibitem{Hofman:2007xp} 
  D.~M.~Hofman and J.~M.~Maldacena,
  ``Reflecting magnons,''
  JHEP {\bf 0711}, 063 (2007)
  [arXiv:0708.2272 [hep-th]].

\bibitem{Berenstein:2009qd} 
  D.~Berenstein and D.~Trancanelli,
  ``S-duality and the giant magnon dispersion relation,''
  arXiv:0904.0444 [hep-th].
  
\bibitem{Das:2000st} 
  S.~R.~Das, A.~Jevicki and S.~D.~Mathur,
  ``Vibration modes of giant gravitons,''
  Phys.\ Rev.\ D {\bf 63}, 024013 (2001)
  [hep-th/0009019].
 
\bibitem{Polchinski:1995mt} 
  J.~Polchinski,
  ``Dirichlet Branes and Ramond-Ramond charges,''
  Phys.\ Rev.\ Lett.\  {\bf 75}, 4724 (1995)
  [hep-th/9510017].
  
 
\bibitem{Grisaru:2000zn} 
  M.~T.~Grisaru, R.~C.~Myers and O.~Tafjord,
  ``SUSY and goliath,''
  JHEP {\bf 0008}, 040 (2000)
  [hep-th/0008015].
 
\bibitem{Hashimoto:2000zp} 
  A.~Hashimoto, S.~Hirano and N.~Itzhaki,
  ``Large branes in AdS and their field theory dual,''
  JHEP {\bf 0008}, 051 (2000)
  [hep-th/0008016].
  
\bibitem{Berenstein:2007wi} 
  D.~Berenstein,
  ``Strings on conifolds from strong coupling dynamics, part I,''
  JHEP {\bf 0804}, 002 (2008)
  [arXiv:0710.2086 [hep-th]].
  
\bibitem{Berenstein:2007kq} 
  D.~E.~Berenstein and S.~A.~Hartnoll,
  ``Strings on conifolds from strong coupling dynamics: Quantitative results,''
  JHEP {\bf 0803}, 072 (2008)
  [arXiv:0711.3026 [hep-th]].


\bibitem{Beisert:2006qh} 
  N.~Beisert,
  ``The Analytic Bethe Ansatz for a Chain with Centrally Extended su(2|2) Symmetry,''
  J.\ Stat.\ Mech.\  {\bf 0701}, P01017 (2007)
  [nlin/0610017 [nlin.SI]].


\bibitem{BFSS} 
  T.~Banks, W.~Fischler, S.~H.~Shenker and L.~Susskind,
  ``M theory as a matrix model: A Conjecture,''
  Phys.\ Rev.\ D {\bf 55}, 5112 (1997)
  [hep-th/9610043].
  
\bibitem{Verlinde:2002ig} 
  H.~L.~Verlinde,
  ``Bits, matrices and 1/N,''
  JHEP {\bf 0312}, 052 (2003)
  [hep-th/0206059].
  
\bibitem{SheikhJabbari:2004ik} 
  M.~M.~Sheikh-Jabbari,
  ``Tiny graviton matrix theory: DLCQ of IIB plane-wave string theory, a conjecture,''
  JHEP {\bf 0409}, 017 (2004)
  [hep-th/0406214].
   
\bibitem{deMelloKoch:2012ck} 
  R.~de Mello Koch and S.~Ramgoolam,
  ``A double coset ansatz for integrability in AdS/CFT,''
  JHEP {\bf 1206}, 083 (2012)
  [arXiv:1204.2153 [hep-th]].
 
\bibitem{Douglas:2001ba} 
  M.~R.~Douglas and N.~A.~Nekrasov,
  ``Noncommutative field theory,''
  Rev.\ Mod.\ Phys.\  {\bf 73}, 977 (2001)
  [hep-th/0106048].
   
    
\end{thebibliography}
\end{document}